\documentclass[12pt]{article}

\newcommand{\om}{\omega}
\newcommand{\al}{\alpha}
\newcommand{\ep}{\epsilon}

\newcommand{\NS}{\mbox{NS}}
\newcommand{\tNS}{\widetilde{\mbox{NS}}}
\newcommand{\R}{\mbox{R}}
\newcommand{\tR}{\widetilde{\mbox{R}}}
\newcommand{\sNS}{\msc{NS}}
\newcommand{\stNS}{\widetilde{\msc{NS}}}
\newcommand{\sR}{\msc{R}}
\newcommand{\stR}{\widetilde{\msc{R}}}

\newcommand{\lb}{\lbrack}
\newcommand{\rb}{\rbrack}

\newcommand{\msc}[1]{\mbox{\scriptsize #1}}
\newcommand{\dsp}{\displaystyle}

\newcommand{\sgn}{\mbox{sgn}}

\newcommand{\bc}{\mbox{{\bf C}}}
\newcommand{\br}{\mbox{{\bf R}}}
\newcommand{\bz}{\mbox{{\bf Z}}}

\newcommand{\n}{\mbox{\bf n}}
\newcommand{\w}{\mbox{\bf w}}
\newcommand{\s}{\mbox{\bf s}}
\newcommand{\p}{\mbox{\bf p}}
\newcommand{\sn}{\msc{\bf n}}
\newcommand{\sw}{\msc{\bf w}}
\renewcommand{\ss}{\msc{\bf s}}
\renewcommand{\sp}{\msc{\bf p}}
\newcommand{\ww}{\widehat{\mbox{\bf w}}}
\newcommand{\ws}{\widehat{\mbox{\bf s}}}
\newcommand{\wsw}{\widehat{\msc{\bf w}}}
\newcommand{\wss}{\widehat{\msc{\bf s}}}
\renewcommand{\a}{\mbox{\bf a}}

\newcommand{\bsz}{\msc{{\bf Z}}}
\newcommand{\bsr}{\msc{{\bf R}}}

\newcommand{\cG}{{\cal G}}

\newcommand{\cN}{{\cal N}}

\newcommand{\cF}{{\cal F}}

\newcommand{\cQ}{{\cal Q}}

\newcommand{\cZ}{{\cal Z}}
\newcommand{\cI}{{\cal I}}
\newcommand{\cK}{{\cal K}}

\newcommand{\X}{\mbox{{\bf X}}}
\newcommand{\Y}{\mbox{{\bf Y}}}
\newcommand{\Z}{\mbox{{\bf Z}}}
\newcommand{\sX}{\msc{{\bf X}}}
\newcommand{\sY}{\msc{{\bf Y}}}
\newcommand{\sZ}{\msc{{\bf Z}}}

\newcommand{\tL}{\tilde{L}}
\newcommand{\tJ}{\tilde{J}}

\newcommand{\tG}{\tilde{G}}

\newcommand{\tq}{\tilde{q}}

\newcommand{\tm}{\tilde{m}}

\newcommand{\hC}{\hat{\cal C}}
\newcommand{\hD}{\hat{\cal D}}

\newcommand{\ket}[1]{{|#1\rangle}}
\newcommand{\bra}[1]{{\langle#1|}}
\newcommand{\dket}[1]{{\left.\left|#1\right\rangle\right\rangle}}
\newcommand{\dbra}[1]{{\left\langle\left\langle#1\right|\right.}}

\newcommand{\dketd}[1]{{\left.\left|#1\right\rangle\right
 \rangle_{\msc{\bf d}}}}
\newcommand{\dketc}[1]{{\left.\left|#1\right\rangle\right
 \rangle_{\msc{\bf c}}}}

\newcommand{\dbrad}[1]{{{}_{\msc{\bf d}}
\left\langle\left\langle#1\right|\right.}}
\newcommand{\dbrac}[1]{{{}_{\msc{\bf c}}
\left\langle\left\langle#1\right|\right.}}
\newcommand{\ketA}[1]{{|#1\rangle}_{\msc{A}}}
\newcommand{\braA}[1]{{}_{\msc{A}}{\langle#1|}}
\newcommand{\ketB}[1]{{|#1\rangle}_{\msc{B}}}
\newcommand{\braB}[1]{{}_{\msc{B}}{\langle#1|}}

\newcommand{\Th}[2]{\Theta_{#1,#2}}
\renewcommand{\th}{{\theta}}

\newcommand{\ch}[2]{\mbox{ch}^{#1}_{#2}}

\newcommand{\chig}{\chi_{\msc{\bf G}}}
\newcommand{\chic}{{\chi_{\msc{\bf c}}}}
\newcommand{\chid}{{\chi_{\msc{\bf d}}}}

\newcommand{\mod}{\mbox{mod}}

\newcommand{\sIm}{\msc{Im}}
\renewcommand{\Re}{\mbox{Re}}

\newcommand{\nn}{\nonumber\\}

\newcommand {\eqn}[1]{(\ref{#1})}

\makeatletter
\@addtoreset{equation}{section}

\makeatother
\begin{document}

\begin{titlepage}
 \
 \renewcommand{\thefootnote}{\fnsymbol{footnote}}
 \font\csc=cmcsc10 scaled\magstep1
 {\baselineskip=14pt
 \rightline{
 \vbox{\hbox{hep-th/0411041}
       \hbox{UT-04-30}
       }}}

 \baselineskip=20pt
\vskip 2cm
 
\begin{center}

{\bf \Large   Conifold Type Singularities, \\

\vskip 5mm

 ${\cal N}=2$ Liouville and $SL(2;\br)/U(1)$  Theories} 

 \vskip 2cm
\noindent{ \large Tohru Eguchi\footnote{eguchi@hep-th.phys.s.u-tokyo.ac.jp}
and Yuji Sugawara\footnote{sugawara@hep-th.phys.s.u-tokyo.ac.jp}} \\
\bigskip

 \vskip .6 truecm
 {\baselineskip=15pt
 {\it Department of Physics,  Faculty of Science, \\
  University of Tokyo \\
  Hongo 7-3-1, Bunkyo-ku, Tokyo 113-0033, Japan}
 }
\end{center}

\bigskip

\begin{abstract}

In this paper we discuss various aspects of 
non-compact models of CFT of  
the type: 
$ \prod_{j=1}^{N_L}\{ \mbox{${\cal N}=2$ Liouville theory}\}_j\otimes
\prod_{i=1}^{N_M} \{\mbox{$\cN=2$ minimal model}\}_i $ and \\
$ \prod_{j=1}^{N_L}\{ \mbox{$SL(2;\br)/U(1)$ supercoset}\}_j\otimes
\prod_{i=1}^{N_M} \{\mbox{$\cN=2$ minimal model}\}_i $. 
These models are related to each other by T-duality.
Such string vacua are expected to describe non-compact Calabi-Yau
compactifications, typically ALE fibrations over (weighted) projective spaces. 
We find that when the Liouville ($SL(2;\br)/U(1)$) theory is 
coupled to minimal models,  there exist 
only $(c,c)$, $(a,a)$ ($(c,a)$, $(a,c)$)-type 
of massless states in CY 3 and 4-folds and 
the theory possesses only complex (K\"{a}hler) structure deformations.
Thus the space-time has the characteristic feature of 
a conifold type singularity whose deformation (resolution) 
is given by the ${\cal N}=2$ Liouville ($SL(2;\br)/U(1)$) theory.

Spectra of compact BPS D-branes determined from the open string sector
are compared with those of massless moduli. 
We compute the open string Witten index and 
determine intersection numbers of vanishing cycles. 
We also study non-BPS branes of the theory that 
are natural extensions of the ``unstable B-branes'' 
of the $SU(2)$ WZW model in \cite{MMS}.

\end{abstract}

\vfill

\setcounter{footnote}{0}
\renewcommand{\thefootnote}{\arabic{footnote}}
\end{titlepage}
\baselineskip 18pt

\section{Introduction}

~

Superstring theories on non-compact curved backgrounds 
are  receiving a great deal of attentions. 
Well-defined description of these string vacua by 
irrational superconformal field theories (SCFT's) is an important 
and challenging problem. 
Recently, considerable progress has been made 
in the study of the $\cN=2$ Liouville theory or
the $SL(2;\br)/U(1)$ Kazama-Suzuki supercoset theory,
based on the method of modular bootstrap and 
the exact description of D-branes in terms of boundary states
\cite{ES-L,ASY,IPT,ES-BH,IKPT,IPT2,ASY2,FNP,NST,Hosomichi,Nakayama2}.
It should be also mentioned that attempts 
of the conformal (boundary) bootstrap in these $\cN=2$ systems 
are given in \cite{ASY,IPT2,ASY2,Hosomichi}.

In this paper we investigate the superstring vacua of the type
$$
\prod_{j=1}^{N_L} \{\mbox{${\cal N}=2$ Liouville}\}_j\otimes
\prod_{i=1}^{N_M}\{\mbox{$\cN=2$ minimal model}\}_i
$$
which in general contain more than one ${\cal N}=2$ Liouville fields 
coupled to ${\cal N}=2$ minimal model. 
A suitable orbifolding procedure is imposed 
as in the Gepner construction \cite{Gepner} in order to ensure the space-time SUSY. 
If one uses the T-duality (mirror symmetry) of ${\cal N}=2$ Liouville 
to $SL(2;\br)/U(1)$ Kazama-Suzuki model \cite{FZZ2,GK,HK1},
the above models are equivalent to
$$
\prod_{j=1}^{N_L} \{\mbox{$SL(2;\br)/U(1)$ supercoset}\}_j
\otimes
\prod_{i=1}^{N_M}\{\mbox{$\cN=2$ minimal model}\}_i.
$$

These models are expected to describe non-compact Calabi-Yau manifolds
where we obtain non-gravitational space-time theories 
due to the Liouville mass gap. 
The earlier studies on such models are given in {\em e.g.}
\cite{MV,GV,OV,ABKS,GKP,GK,Pelc,ES1,Lerche,LLS,ES2,HK2,ncCY-others}.
Many related topics and a detailed list of literature 
are found in a review paper \cite{Nakayama}.

A basic idea in Gepner construction is the orbifolding  
with respect to the $U(1)_R$-charge of $\cN=2$ 
superconformal algebra (SCA). One way to impose charge-integrality  
is to consider spectral-flow orbits as in  \cite{EOTY}: by 
using flow-invariant orbits we can systematically construct 
conformal blocks of the theory. 
In our previous paper \cite{ES-L}, we have considered
$\cN=2$ Liouville theory with rational central charges and 
introduced extended characters which are defined by  an infinite sum
over spectral flows of irreducible ${\cal N}=2$ characters. 
We have shown that  
\begin{itemize}
 \item Extended characters have discrete 
and finite spectra of $U(1)_R$-charges, although
  they may have continuous spectra of conformal weights.
 \item They are closed under modular transformations.
\end{itemize}
We have also noticed that
these characters naturally appear in the 
torus partition functions of $SL(2;\br)/U(1)$ Kazama-Suzuki models
\cite{ES-BH} (see also \cite{IKPT}), which are 
T-dual to the $\cN=2$ Liouville theories. 
In this paper we use extended characters together with 
irreducible characters of minimal models 
as basic building blocks of our construction. 

In the following we especially study models 
with $N_M=1$ and $1\leq N_L \leq 3$, 
which are interpreted geometrically as ALE fibrations 
over (weighted) projective spaces \cite{Lerche,HK2}.
We find a very 
interesting aspect of massless spectrum in these models:
in the case of ${\cal N}=2$ Liouville theory coupled to minimal models  
there exist only $(c,c)$ or $(a,a)$-type massless states 
in the CY 3 and 4-folds and no  
$(c,a)$ or $(a,c)$ massless states appear. Thus the 
theory possesses only complex structure deformations 
and no deformations of K\"ahler structure.
On the other hand, if we use the $SL(2;\br)/U(1)$ description, 
the theory possesses only $(c,a)$ and $(a,c)$-type massless states 
and the moduli of K\"{a}hler structure deformations.
Thus the space-time has the characteristic feature 
of a conifold type singularity which is deformed (resolved) by
the ${\cal N}=2$ Liouville ($SL(2;\br)/U(1)$) theory.

In the case of models describing non-compact K3 surfaces 
or smoothed ADE type singularity, 
on the other hand,
the same number of $(a,c),(c,a)$ and $(c,c),(a,a)$ states 
appear in accord with the
 ${\cal N}=4$ world-sheet supersymmetry.

This paper is organized as follows: in section 2
we present a brief review on the irreducible and 
extended characters in the $SL(2;\br)/U(1)$ Kazama-Suzuki models 
following \cite{ES-BH}. In section 3 we study the closed string 
sector of our non-compact models. We analyze
the torus partition functions, elliptic genera and the massless 
spectra of closed string states. We study models with 
$N_M=1$ and $1\leq N_L \leq 3$, and find the interesting 
characteristics of their massless spectra as mentioned above.


In section 4, we study the open string sector of our models. We 
focus on the BPS compact branes and evaluate the cylinder amplitudes. 
We compare the spectra of BPS compact branes with 
those of massless moduli determined in section 3. We find that 
some of the BPS branes (cycles) are not
associated with massless moduli as noticed previously 
in the case of singular CY manifolds \cite{GVW,GKP,Pelc}.
We also derive the general formula of open string Witten 
indices and prove the conjecture of \cite{Lerche}. 
We further construct boundary states for a class 
of non-BPS D-branes which are extensions of 
``unstable B-branes'' in the
$SU(2)$-WZW model \cite{MMS}.  
Contrary to the flat case, non-BPS branes including 
RR-components (but with vanishing RR-charges)
also exist. This type of branes could be identified 
with the ones studied recently in \cite{Kutasov2} using
the DBI action. 
Section 5 is devoted to a summary and discussions. 
We present in Appendix E some consistency checks 
of our modular transformation formulas with the known 
results about the higher level Appell functions \cite{Pol,STT}.

In the following we mainly use the language of 
$SL(2;\br)/U(1)$ supercoset theory 
rather than the ${\cal N}=2$ Liouville theory for the sake of convenience.
However, later in section 3 we identify results of CFT analyses as 
describing the deformed geometries based on  ${\cal N}=2$ Liouville theory.



~

\newpage

\section{Preliminaries}

~

We start with a brief review on the conformal blocks 
and their modular properties of $SL(2;\br)/U(1)$ Kazama-Suzuki model. 
More complete arguments are given in \cite{ES-BH}
(see also \cite{IKPT}).

~

\subsection{Branching Functions in $SL(2;\br)/U(1)$ Kazama-Suzuki model}

~

The Kazama-Suzuki model for $SL(2;\br)/U(1)$ 
is defined as the coset CFT
\begin{eqnarray}
\frac{SL(2;\br)_{\kappa}\times SO(2)_1}{U(1)_{-(\kappa-2)}}~,
\end{eqnarray}
which is an $\cN=2$ SCFT 
with $\hat{c}(\equiv c/3)=1+2/k$, ($k\equiv \kappa-2$).
The coset characters are defined by the following branching relation
\begin{eqnarray}
\chi_{\xi}\left(\tau,\frac{2}{k}z+w\right)\frac{
\th_3\left(\tau,\frac{k+2}{k}z+w\right)}{\eta(\tau)}
=\sum_{m}\chi^{(\sNS)}_{\xi,m}(\tau,z)
\frac{q^{-\frac{m^2}{k}}e^{2\pi i mw}}{\eta(\tau)}~.
\label{branching 0}
\end{eqnarray}
where $\xi$ labels irreducible representations of 
$\widehat{SL}(2;\br)_{\kappa}$ and $\chi_{\xi}(\tau,u)$ denotes its 
character. 
We can identify the branching functions 
$\chi^{(\sNS)}_{\xi,m}(\tau,z)$ with the irreducible characters of
$\cN=2$ SCA as follows:
\begin{itemize}
 \item {\bf for the continuous series $\xi =\hC_{p,m}$ : }
\begin{eqnarray}
\chic^{(\sNS)}_{\,p,m}(\tau,z) &=& 
q^{\frac{p^2}{2}+\frac{m^2}{k}}\, 
e^{2\pi i \frac{2m}{k}z} \,\frac{\th_3(\tau,z)}{\eta(\tau)^3}~,
\label{branching C}
\end{eqnarray}
which are the massive characters of $\cN=2$ SCA.
The highest-weight state has the quantum numbers, conformal dimension and
$U(1)$-charge as
\begin{eqnarray}
&& h= \frac{p^2}{2}+\frac{1}{4k}+\frac{m^2}{k}~, ~~~ Q=\frac{2m}{k}~.
\end{eqnarray}
 \item {\bf for the discrete series $\xi=\hD^+_j$ : }
\begin{eqnarray}
\chid_{\,j,m=j+n}^{(\sNS)}(\tau,z) 
&=& \frac{q^{\frac{j+n^2+2nj}{k}-\frac{1}{4k}} 
e^{2\pi i \frac{2(j+n)}{k}z}}{1+e^{2\pi i z}q^{n+1/2}}\,
\frac{\th_3(\tau,z)}{\eta(\tau)^3}~, ~~~ ({}^{\forall}n\in\bz)~,
\label{branching D}
\end{eqnarray}
which are the $n$-step 
spectral flow of massless matter characters. The $n$-step flow 
is generated by an operator $U_n \equiv e^{in\Phi}$ 
where $\Phi$ denotes the zero mode of a scalar field 
of the ${\cal N}=2$ $U(1)$ current $J\equiv i \partial \Phi$.
The unitarity requires the condition \cite{BFK,DPL}
\begin{eqnarray}
0<j<\frac{\kappa}{2}\left(\equiv \frac{k+2}{2}\right)~.
\label{unitarity bound}
\end{eqnarray}
The highest-weight states have the following quantum numbers;
\begin{eqnarray}
&&h= \frac{2j\left(n+\frac{1}{2}\right)+n^2}{k}~,~~~ Q= \frac{2(j+n)}{k}~,~~~
(n\geq 0) ~,
\label{vacuum 1} \\
&& h= \frac{-\left(k-2j\right)\left(n+\frac{1}{2}\right)+n^2}{k}~,
~~~ Q= \frac{2(j+n)}{k}-1~,~~~ (n<0)~.
\label{vacuum 2}
\end{eqnarray}
They  are given explicitly by $(j^+_0)^n\ket{j,j}\otimes \ket{0}_{\psi}$
($n\geq 0$), $(j^-_{-1})^{|n|-1}\ket{j,j}\otimes \psi^-_{-1/2}\ket{0}_{\psi}$
($n<0$) respectively 
(here $|j,j\rangle$ denotes 
the lowest weight state of bosonic
$SL(2;\br)$ algebra and $\ket{0}_{\psi}$ denotes the fermion Fock vacuum).
The highest weight representations $\hD^-_j$ merely yield the same type of 
characters and we need not take account of them. 
\item {\bf for the identity representation $\xi= \mbox{id}$ :}
\begin{eqnarray}
\chi^{(\sNS)}_{0,n}(\tau,z) 
&=& 
q^{-\frac{1}{4k}} \frac{(1-q)q^{\frac{n^2}{k}+n-\frac{1}{2}}
  e^{2\pi i \left(\frac{2n}{k}+1\right)z}}
  {\left(1+e^{2\pi i z}q^{n+1/2}\right)\left(1+e^{2\pi i z}q^{n-1/2}\right)}\,
  \frac{\th_3(\tau,z)}{\eta(\tau)^3}~,
\label{branching Id}
\end{eqnarray}
which are the spectrally flowed graviton characters. 
The vacuum states are summarized as
\begin{description}
 \item [$n=0$] : the vacuum is $\ket{0,0}\otimes \ket{0}_{\psi}$ with 
$h=Q=0$.
 \item [$n\geq 1$] : the vacuum is $(j^+_{-1})^{n-1}\ket{0,0} \otimes 
\psi^+_{-1/2}\ket{0}_{\psi}$, which has the quantum numbers 
 \begin{eqnarray}
  h= \frac{n^2}{k}+n-\frac{1}{2}~,~~~ Q= \frac{2n}{k}+1~. 
 \end{eqnarray}
 \item [$n\leq -1$] : the vacuum is $(j^-_{-1})^{|n|-1}\ket{0,0} \otimes 
\psi^-_{-1/2}\ket{0}_{\psi}$, which has the quantum numbers
 \begin{eqnarray}
  h= \frac{n^2}{k}-n-\frac{1}{2}~,~~~ Q= \frac{2n}{k}-1~. 
 \end{eqnarray}
\end{description}
\end{itemize}

~


\subsection{Extended Characters}

~

From now on, we shall concentrate on models with a rational level 
$k=N/K$ ($N,K\in \bz_{>0}$).
We define the extended characters by taking the mod $N$ spectral flow
sums of irreducible characters.
In the following definitions, $m$ is assumed to be an integral
parameter within the range  $\dsp -NK \leq m < NK$.
\begin{itemize}
 \item {\bf continuous representation (`extended massive character') : }

We define
\begin{eqnarray}
\chic^{(\sNS)}(p,m;\tau,z)& \equiv& \sum_{n\in N\bsz}\,
 \chic^{(\sNS)}_{\,p,m/2K +n} (\tau,z) 
\equiv  q^{\frac{p^2}{2}} \Th{m}{NK}\left(\tau,\frac{2z}{N}\right)\,
  \frac{\th_3(\tau,z)}{\eta(\tau)^3}~,
\label{chi c}
\end{eqnarray}
which has the highest-weight state with 
\begin{eqnarray}
h=\frac{p^2}{2}+\frac{m^2+K^2}{4NK}~,~~~ Q=\frac{m}{N}~.
\label{vacua chi c}
\end{eqnarray}
\item {\bf discrete representation (`extended massless matter character'): }

We define (with the reparameterization $j\equiv s/(2K)$, $s\in \bz$)
\begin{eqnarray}
\chid^{(\sNS)}(s,m;\tau,z) \equiv
\left\{
\begin{array}{ll}
 \sum_{n\in N\bsz}\,\chid^{(\sNS)}_{\,\frac{s}{2K},\frac{m}{2K}+n}(\tau,z)&
 ~~ m \equiv s~(\mod~2K) \\
 0 &~~ m\not\equiv s ~(\mod~2K)
\end{array}
\right.
\label{chi d}
\end{eqnarray}
where the unitarity condition \eqn{unitarity bound} imposes
\begin{eqnarray}
1\leq s \leq N+2K-1~,~~~ (s\in \bz)~.
\label{range s unitarity}
\end{eqnarray}
The vacuum vectors for $\chid^{(\sNS)}(s,m=s+2Kr)$,  
($\dsp -\frac{N}{2}\leq r < \frac{N}{2}$) are characterized by 
\begin{eqnarray}
&& h=\frac{Kr^2+\left(r+\frac{1}{2}\right)s}{N}~, ~~
   Q = \frac{s+2Kr}{N}~, ~~~ \left(0\leq r < \frac{N}{2}\right)\nn
&& h=\frac{Kr^2-\left(r+\frac{1}{2}\right)(N-s)}{N}~, ~~
   Q = \frac{s-N+2Kr}{N}~, ~~~ \left(-\frac{N}{2}\leq r <0\right)~.
\label{vacua chi d}
\end{eqnarray}

\item{\bf identity representation  (`extended graviton character'): }

We define
\begin{eqnarray}
\chi_0^{(\sNS)}(m;\tau,z) \equiv
\left\{
\begin{array}{ll}
 \sum_{n\in N\bsz}\,\chi^{(\sNS)}_{0,\frac{m}{2K}+n}(\tau,z)&
 ~~ m \in 2K\bz \\
 0 &~~ m\not\in  2K\bz 
\end{array}
\right.
\label{chi 0}
\end{eqnarray}
The vacua for $\chi_0^{(\sNS)}(m=2Kr;\tau,z)$, 
$\dsp \left(-\frac{N}{2}\leq r < \frac{N}{2}\right)$
are given as
\begin{eqnarray}
&& h=Q=0~, ~~~ (r=0)~, \nn
&& h=\frac{Kr^2}{N}+|r| -\frac{1}{2}~,~~ Q=\frac{2Kr}{N}+\mbox{sgn}(r)~,
~~~ (r\neq 0)~.
\label{vacua chi 0}
\end{eqnarray}
\end{itemize}


The extended characters of other spin structures are defined by the 
1/2-spectral flow;
\begin{eqnarray}
&& \chi_*^{(\stNS)}(*,m;\tau,z) \equiv e^{-i\pi \frac{m}{N}}\,
\chi_*^{(\sNS)}\left(*,m;\tau,z+\frac{1}{2}\right)~, \nn
&& \chi^{(\sR)}_{*}(*,m+K;\tau,z) \equiv 
  q^{\frac{\hat{c}}{8}} e^{i\pi \hat{c}z}\,
 \chi^{(\sNS)}_{*}\left(*,m;\tau,z+\frac{\tau}{2}\right)~, \nn
&& \chi^{(\stR)}_{*}(*,m+K;\tau,z) \equiv 
  e^{-i\pi \frac{m}{N}} q^{\frac{\hat{c}}{8}} e^{i\pi  \hat{c} z}\,
 \chi^{(\sNS)}_{*}\left(*,m;\tau,z+\frac{\tau}{2}+\frac{1}{2}\right)~.
\label{extended os}  
\end{eqnarray}
Note that extended characters of
discrete and identity representations in R and $\tR$-sectors
take non-zero values 
only if $m\equiv s-K~(\mod\, 2K)$, $m \in K(2\bz+1)$, respectively. 
The quantum numbers 
of the NS and R vacua are related by 
\begin{eqnarray}
 && h^{(\sR)}(*,m+K) = h^{(\sNS)}(*,m+K) + \frac{1}{8} \equiv 
    h^{(\sNS)}(*,m)+\frac{1}{2}Q^{(\sNS)}(m)+\frac{\hat{c}}{8}~, \nn
 && Q^{(\sR)}(m+K) = Q^{(\sNS)}(m+K) +\frac{1}{2} \equiv
    Q^{(\sNS)}(m)+ \frac{\hat{c}}{2}~,
\label{relation NS R}
\end{eqnarray}
where $h^{(\sNS)}(*,m)$, $Q^{(\sNS)}(m)$ are those 
given in \eqn{vacua chi c}, \eqn{vacua chi d} and \eqn{vacua chi 0}. 
Useful properties of the extended characters \eqn{chi c}, \eqn{chi d}
and \eqn{chi 0} are summarized in Appendix C.


The non-compactness of $SL(2;\br)/U(1)$ model 
leads to an IR divergence in the torus partition function.
One may introduce 
an IR cut-off $\epsilon$ and then the regularized partition function
contains a piece which consists of continuous representations 
and also a piece consisting of discrete representations \cite{HPT}. 
Since the continuous representations describe string modes 
propagating in the bulk, their contributions are proportional 
to the volume of target space $V(\epsilon)$,
while the discrete representations describe 
localized string states and their contributions 
are volume independent.
 
Leading terms in the infinite volume limit 
are given by continuous representations
\cite{ES-BH};
\begin{eqnarray}
&&\lim_{\ep\,\rightarrow\,+0} \frac{Z(\tau;\ep)}{V(\ep)} \propto
\frac{1}{2}\sum_{\sigma}\, 
\int_0^{\infty} dp\,
 \sum_{w\in 2K}\,\sum_{n\in N}\, 
\chic^{(\sigma)}(p, Kn+Nw;\tau,0)
\chic^{(\sigma)}(p, -Kn+Nw;-\bar{\tau},0)~. \nn
&&
\label{MI part fn}
\end{eqnarray} 
Here $\sigma$ denotes the spin structure and the above partition function 
is modular invariant. 
The quantum numbers $n$, $w$ are identified with the KK momenta and 
winding modes along the circle of the Euclidean cigar geometry 
with an asymptotic radius $\sqrt{2k} \equiv \sqrt{2N/K}$ of  
the $SL(2;\br)/U(1)$-coset theory \cite{2DBH}.

If one considers the $\tR$-part of the partition function, 
contributions of continuous representations drop out  and only the 
discrete representations survive.  
They give rise to a volume-independent finite result. 
This is nothing but the Witten index;
\begin{eqnarray}
&& Z^{(\stR)}(\tau) = 
\sum_{s=K}^{N+K} \,\sum_{w\in\bsz_{2K}}\, \sum_{n\in \bsz_N}\,
a(s) \,  \chid^{(\stR)}(s, Kn+Nw;\tau,0)
\chid^{(\stR)}(s, -Kn+Nw;-\bar{\tau},0)~, \nn
&& \hspace{3cm}
a(s) \equiv
\left\{
\begin{array}{ll}
 1& ~~K+1 \leq s \leq N+K-1 \\
 \frac{1}{2} & ~~ s=K,N+K
\end{array}
\right. ~~.
\label{discrete part fn}
\end{eqnarray} 
It is important to note that the quantum number $s$ runs over 
the range \cite{ES-BH,IKPT} ;
\begin{eqnarray}
 K \leq s \leq N+K~,
\label{range s}
\end{eqnarray}
which is strictly smaller than \eqn{range s unitarity} if $K\neq 1$
(see also \cite{HPT}).
This range is consistent with the modular transformation formulas 
\eqn{S discrete}, \eqn{S graviton}.

~


\section{Non-compact Cosets Coupled to Minimal Models}

~

Now, we work on the main subject in this paper. 
Let us study the superconformal system defined as 
\begin{eqnarray}
\left\lb
L_{N_1,K_1} \otimes \cdots \otimes L_{N_{N_L}, K_{N_L}} \otimes 
M_{k_1} \otimes \cdots \otimes M_{k_{N_M}} 
\right\rb_{\msc{$U(1)$-projection}}~,
\label{nc gepner}
\end{eqnarray}
where $L_{N,K}$ denotes the $SL(2;\br)/U(1)$
Kazama-Suzuki model with $k=N/K$ ($\hat{c}= 1+2K/N$) and 
$M_k$ denotes the level $k$ $\cN=2$ minimal model with 
$\hat{c}=k/(k+2)$. 
We impose the criticality condition for the case of 
target manifold with (complex) dimension $\n$
\begin{eqnarray}
\sum_{i=1}^{N_M} \frac{k_i}{k_i+2} + \sum_{j=1}^{N_L} 
\left(1+\frac{2K_j}{N_j}\right) = \n~, ~~~ \n=2,3,4~.
\label{criticality}
\end{eqnarray}
Since $\hat{c}>1$ for each $SL(2;\br)/U(1)$-sector, 
it is obvious that $N_L \leq \n-1$ holds.
We expect that the $U(1)$-charge projection 
yields consistent superstring vacua describing non-compact 
$CY_{\sn}$ compactifications with $d$-dimensional 
Minkowski space $(d=10-2\n)$. 
Note that the periodicity of extended characters 
depends on the choice of $N_j$, $K_j$, not only on the ratio
$N_j/K_j$. 
We shall thus adopt the notation $L_{N_j,K_j}$ to indicate 
which extended characters are used, 
although only the ratio $N_j/K_j$ parameterizes  the $SL(2;\br)/U(1)$
supercoset.
For simplicity we here assume  that each pair 
$N_j$, $K_j$ is  relatively prime for every $j=1,\ldots, N_L$. 
We set 
\begin{eqnarray}
N \equiv \mbox{L.C.M.}\left\{k_i+2, \, N_j\right\}~,~~~i=1,
\cdots,N_M, ~~~ j=1,\cdots,N_L~,
\end{eqnarray}
and then the required $U(1)$-projection is reduced to 
the $\bz_N$-orbifoldization. 
We introduce the notations 
\begin{eqnarray}
\mu_i,\nu_j \in \bz_{>0}~,~~ N=\mu_i (k_i+2)= \nu_j N_j~,
\label{mu nu}
\end{eqnarray}
for later convenience. 

~

\subsection{Toroidal Partition Functions : Continuous Part of 
Closed String Spectra}

~

We first analyse the closed string sector.  Only the continuous part of 
closed string spectrum contributes to the modular invariant partition 
function (per unit volume), and should be interpreted as the propagating 
modes in the non-compact Calabi-Yau space. 
More detailed argument has been given in \cite{ES-BH} for 
the case $N_L=1$. 
Let us start by assuming 
\begin{itemize}
 \item diagonal modular invariance in each $M_{k_i}$-sector,
 \item the partition function \eqn{MI part fn} for each $L_{N_j,K_j}$-sector,
\end{itemize}
before performing the $\bz_N$-orbifoldization. 
As in the standard treatment of orbifold, the $\bz_N$-projection
must be accompanied by the twisted sectors generated by the 
spectral flows.
The integral spectral flows act on each character as the shifts 
of the angular variable; $z\,\mapsto\,z+a\tau+b$ ($a,b \in \bz_N$),
and thus 
the relevant conformal blocks are defined as the flow invariant
orbits \cite{EOTY};
\begin{eqnarray}
&&\cF^{(\sNS)}_{I,\sp,\sw}(\tau,z) \equiv 
\frac{1}{N}\sum_{a,b\in \bsz_{N}}\, q^{\frac{\sn}{2}a^2}e^{2\pi i \sn a z}\,
\prod_{i=1}^{N_M} 
\ch{(\sNS)}{\ell_i,m_i}
   (\tau,z+a\tau+b) \, \nn
&& \hspace{4cm}\times
 \prod_{j=1}^{N_L} 
\chic^{(\sNS)}(p_j,K_jn_j+N_jw_j;\tau,z+a\tau+b)~, \nn
&&\tilde{\cF}^{(\sNS)}_{I,\sp,\sw}(-\bar{\tau},\bar{z}) \equiv 
\frac{1}{N}\sum_{a,b\in \bsz_{N}}\, \bar{q}^{\frac{\sn}{2}a^2}
e^{2\pi i \sn a \bar{z}}\,
\prod_{i=1}^{N_M} 
\ch{(\sNS)}{\ell_i,m_i}
   (-\bar{\tau},\bar{z}+a\bar{\tau}+b) \, \label{cF cont} \\
&& \hspace{4cm}\times
 \prod_{j=1}^{N_L} 
\chic^{(\sNS)}(p_j,-K_jn_j+N_jw_j;-\bar{\tau},
-\bar{z}-a\bar{\tau}-b) \equiv 
\cF^{(\sNS)}_{I,\sp,-\sw}(-\bar{\tau},\bar{z})~,
 \nn
&& I= \left((\ell_1,m_1),\ldots, (\ell_{N_M},m_{N_M}), 
n_1,\ldots, n_{N_L}\right)~, ~~~ \p=(p_1,\ldots,p_{N_L})~,~~~
\w=(w_1,\ldots,w_{N_L})~,
\nonumber
\end{eqnarray}
where $\ch{(\sNS)}{\ell_i,m_i}(\tau,z)$ denotes the character of
the minimal
model $M_{k_i}$ and 
$\chic^{(\sNS)}(p_j,K_jn_j+N_jw_j;\tau,z)$ is the extended 
character \eqn{chi c} of the $SL(2;\br)/U(1)$ theory $L_{N_j,K_j}$.\footnote
{In the right-moving sector of $SL(2;\br)/U(1)$ theory we have chosen an angular dependence $-\bar{z}$. This is in order to bring our convention of quantum numbers $n_j,w_j$ consistent with those
given by the cigar geometry of 2-dimensional black hole.}
The summation $\dsp \frac{1}{N}\sum_{b \in \bsz_N} \, *$ imposes 
the constraint on the $U(1)$-charge
\begin{eqnarray}
\sum_{i=1}^{N_M}\,\frac{m_i}{k_i+2} + 
\sum_{j=1}^{N_L}\,\frac{K_j n_j}{N_j} \in \bz~,
\label{U(1) charge constraint}
\end{eqnarray}
and we automatically have $\cF_{*}^{(\sNS)} \equiv 0$ unless 
\eqn{U(1) charge constraint} is satisfied.
The conformal blocks of other spin structures are defined by 
the $1/2$-spectral flows;
\begin{eqnarray}
&&\cF^{(\stNS)}_{I,\sp,\sw}(\tau,z) \equiv
\cF^{(\sNS)}_{I,\sp,\sw}\left(\tau,z+\frac{1}{2}\right) , \,\,\,
\cF^{(\sR)}_{I,\sp,\sw}(\tau,z) \equiv
q^{\frac{\sn}{8}} e^{i\pi \sn z}
\cF^{(\sNS)}_{I,\sp,\sw}\left(\tau,z+\frac{\tau}{2}\right) ~, \nn 
&&\cF^{(\stR)}_{I,\sp,\sw}(\tau,z) \equiv
q^{\frac{\sn}{8}} e^{i\pi \sn z}
\cF^{(\sNS)}_{I,\sp,\sw}\left(\tau,z+\frac{\tau}{2}+\frac{1}{2}\right) ~.
\label{cF cont os}
\end{eqnarray}
By construction the conformal blocks $\cF^{(\sigma)}_{I,\sp,\sw}$ 
have the following symmetry (${}^{\forall}a,{}^{\forall}b\in \bz$)
\begin{eqnarray}
q^{\frac{\sn}{2}a^2}e^{2\pi i \sn a z}
\cF^{(\sNS)}_{I,\sp,\sw}(\tau,z+a\tau+b)
&=& \cF^{(\sNS)}_{I,\sp,\sw}(\tau,z)~,  \nn
q^{\frac{\sn}{2}a^2}e^{2\pi i \sn a z}
\cF^{(\stNS)}_{I,\sp,\sw}(\tau,z+a\tau+b)
&=& (-1)^{\sn a} \cF^{(\stNS)}_{I,\sp,\sw}(\tau,z) ~, \nn
q^{\frac{\sn}{2}a^2}e^{2\pi i \sn a z}
\cF^{(\sR)}_{I,\sp,\sw}(\tau,z+a\tau+b)
&=& (-1)^{\sn b} \cF^{(\sR)}_{I,\sp,\sw}(\tau,z) ~, \nn
q^{\frac{\sn}{2}a^2}e^{2\pi i \sn a z}
\cF^{(\stR)}_{I,\sp,\sw}(\tau,z+a\tau+b)
&=& (-1)^{\sn (a+b)} \cF^{(\stR)}_{I,\sp,\sw}(\tau,z) ~, 
\label{s f inv}
\end{eqnarray}
Taking the diagonal modular invariance for the spin structures, 
we obtain the partition function (as a non-linear 
$\sigma$-model) 
\begin{eqnarray}
&&Z(\tau,z) = e^{-2\pi \sn \frac{\left(\sIm \, z\right)^2}{\tau_2}}\,
\frac{1}{2N}\sum_{\sigma}\,\sum_{I,\sw}\,\int d^{N_L}\p\,
\cF^{(\sigma)}_{I,\sp,\sw}(\tau,z)
\cF^{(\sigma)}_{I,\sp,-\sw}(-\bar{\tau},\bar{z})~.
\label{MI part fn sigma model}
\end{eqnarray}
The overall factor $1/N$ is necessary to avoid the overcounting of 
states. 
One can easily check the modular invariance of this partition function 
using the modular properties
of $L_{N_j,K_j}$, $M_{k_i}$ given in appendices. 
A crucial point is 
the fact that the sum over the spectral flow 
$z\,\mapsto\,z+a\tau+b$ in \eqn{cF cont}  
preserves modular invariance.

Incorporating the flat space-time $\br^{d-1,1}$ 
(with $\dsp \frac{d}{2}+\n=5$), we also obtain the supersymmetric 
conformal blocks of superstring vacua
\begin{eqnarray}
\frac{1}{\tau_2^{\frac{d-2}{4}}\eta(\tau)^{d-2}}\,\sum_{\sigma}\,
  \ep(\sigma)\, 
  \left(\frac{\th_{\lb \sigma \rb}(\tau)}{\eta(\tau)}\right)^{\frac{d-2}{2}}
  \, \cF^{(\sigma)}_{I,\sp,\sw}(\tau,0)~,
\label{SUSY conf block}
\end{eqnarray}
where $\th_{\lb \sigma \rb}$ denotes $\th_3$, $\th_4$, $\th_2$,
$i\th_1$ for $\sigma = \NS, \tNS, \R, \tR$ respectively, and we 
set $\ep(\NS)=\ep(\tR)=+1$, $\ep(\tNS)=\ep(\R)=-1$. 
The conformal blocks \eqn{SUSY conf block} actually vanish for 
arbitrary $\tau$ \cite{HS}. 

One can choose a large variety of modular invariants as consistent 
conformal theories. For example, one may take general modular 
invariants of the types given in \cite{GQ} with respect to
$\w \in \bz_{2K_1}\times \cdots \times \bz_{2K_{N_L}}$.

It turns out that some of the familiar non-compact Calabi-Yau spaces 
do not correspond to the simplest choice of modular invariance 
\eqn{MI part fn sigma model}
and we have to use a somewhat more non-trivial form of modular invariant.
A typical example 
showing such peculiar feature is 
the singular $CY_3$ of 
$A_{n-1}$-type ($CY_3(A_{n-1})$) \cite{GVW,GKP}. 
The conformal blocks for this model presented in 
\cite{ES1} are  written (with suitable change of notations) as 
\begin{eqnarray}
&& \hspace{-1cm}
\cF^{(\sNS)}_{\ell,w}(\tau,z) = \sum_{m\in \bsz_{4n}}\, 
\ch{(\sNS)}{\ell,m}(\tau,z) \, \frac{\Th{-(n+2)m+2nw}{2n(n+2)}
\left(\tau,\frac{z}{n}\right)}{\eta(\tau)}~, 
~~~ \ell +2w \in 2\bz~, ~~(w \in \frac{1}{2}\bz_{4(n+2)})~, \nn
&&
\label{ES conf-block CY3} 
\end{eqnarray}
where we omitted the factor depending on the `Liouville momentum' $p$.
These are identified with the branching functions of the coset CFT: ~ 
$\dsp \frac{SU(2)_{n-2}\times SO(4)_1}{U(1)_{n+2}}$.
At first glance, \eqn{ES conf-block CY3} seems to fit with the formulas 
\eqn{cF cont} with $N=2n$, $K=n+2$. 
However, {\em half-integral} values of $w$ are now allowed with the constraint
\begin{eqnarray}
m+2w \in 2\bz~.
\label{cond m w CY3}
\end{eqnarray}
This condition may be interpreted 
as some kind of  orbifoldization, and makes it possible 
to pair each of  the primary states of the minimal model $M_{n-2}$ to
those of $L_{2n,n+2}$  so that it yields a physical state with 
an integral $U(1)$-charge. As we shall see later, 
we obtain the expected spectrum of massless states 
as the singular $CY_3$ of $A_{n-1}$-type and 
the correct open string Witten indices under this condition 
\eqn{cond m w CY3}. 
We denote the $SL(2;\br)/U(1)$-sector defined this way as 
$L'_{2n,n+2}$ from now on.

A similar example which we will later study 
is a model with two Liouville fields 
$N_L=2$ and $N_M=1$;
\begin{eqnarray}
\hat{c}=4~,~~~k_1=n-2~,~~~ N_1=N_2=4n~,~~~K_1=K_2=n+2~.
\end{eqnarray}
It is possible to show that the following conformal blocks give the consistent 
superstring vacuum;
\begin{eqnarray}
&& \hspace{-1cm}
\cF^{(\sNS)}_{\ell,m,m_j,w_j, p_j} (\tau,z)
= \frac{1}{4n}\sum_{a\in \bsz_{4n}}\, \ch{(\sNS)}{\ell,m-2a}(\tau,z)
\,
\prod_{j=1,2} \left\{
\chic^{(\sNS)}_{(4n,n+2)}(p_j, (n+2)m_j+4nw_j+2(n+2)a;\tau,z) \right. \nn
&& \hspace{2cm} \left. +
\chic^{(\sNS)}_{(4n,n+2)}(p_j, (n+2)(m_j+4n)+4nw_j+2(n+2)a;\tau,z)
\right\}~ \nn
&&  
\equiv  \frac{1}{2n} \sum_{a\in \bsz_{2n}}\, \ch{(\sNS)}{\ell,m-2a}(\tau,z)
\,\prod_{j=1,2} 
\chic^{(\sNS)}_{(2n, \frac{n+2}{2})}
\left(p_j, \frac{n+2}{2} m_j+2nw_j+(n+2)a;\tau,z\right) 
~,~~\mbox{(if $n$ is even)}~, \nn  
&&  
\label{cF CY3 fiber}
\end{eqnarray}
with 
\begin{eqnarray}
m \in \bz_{2n}~,~~ m_j \in \bz_{4n}~, ~~ w_j \in \frac{1}{4}\bz_{4(n+2)}~,
~~ m_j+4w_j \in 2\bz~, ~~ \sum_{j=1,2}(m_j+4w_j) \in 4\bz~,
\label{m CY3 fiber}
\end{eqnarray}
and the $U(1)$-charge condition 
\begin{eqnarray}
2m+m_1+m_2 \in 2n\bz~.
\label{U(1) cond CY3 fiber}
\end{eqnarray}
We here use the notation $\chic^{(*)}_{(N,K)}(*,*;\tau,z)$
with the parameters $N$, $K$ written explicitly.
By careful calculations it is possible to show
that the conformal blocks  \eqn{cF CY3 fiber} 
are in fact closed under modular transformations in a manner 
consistent with the non-trivial restrictions  
\eqn{m CY3 fiber}, \eqn{U(1) cond CY3 fiber}.
The coefficients of S-transformation include the factors 
$$
\frac{1}{\sqrt{4n}} e^{-2\pi i \frac{m_jm_j'}{4n}} \cdot
\frac{1}{\sqrt{4(n+2)}}e^{2\pi i \frac{4 w_j w_j'}{n+2}}
$$
in each $L'_{4n,n+2}$-sector, and we can construct modular invariants 
in the standard way.
Hence this model yields a consistent string vacuum with 
the choice of spectrum \eqn{m CY3 fiber}. 
We denote the system defined this way  as $L'_{4n,n+2}$.
All the primaries in $M_{n-2}$ can again find partners in 
two $L'_{4n,n+2}$-sectors. 
We will later identify this vacuum 
as a non-compact $CY_4$ with a singular $CY_3$ fibered over $\bc P^1$.

~

\subsection{Elliptic Genera : Discrete Part of Closed String Spectra}

~

Let us turn to the discrete spectrum in the closed 
string Hilbert space. It describes localized string excitations 
and is of basic importance since massless states appear in this sector. 
A useful quantity that captures the BPS states
is the elliptic genus, and we try to evaluate it 
for general models \eqn{nc gepner}. It is basically a generalization
of the analysis given in \cite{ES-BH} to the case $N_L \geq 1$.
We consider flow invariant orbits consisting of
the discrete characters \eqn{chi d} in place of \eqn{cF cont};
\begin{eqnarray}
&& \cG^{(\sNS)}_{I,\ss,\sw}(\tau,z) \equiv 
\frac{1}{N}\sum_{a,b\in \bsz_{N}}\, q^{\frac{\sn}{2}a^2}e^{2\pi i \sn a z}\,
\prod_{i=1}^{N_M} 
\ch{(\sNS)}{\ell_i,m_i}
   (\tau,z+a\tau+b) \, \nn
&& \hspace{4cm}\times
 \prod_{j=1}^{N_L} 
\chid^{(\sNS)}(s_j,K_jn_j+N_jw_j;\tau,z+a\tau+b)~, 
\label{cG} \\
&&\tilde{\cG}^{(\sNS)}_{I,\ss,\sw}(-\bar{\tau},\bar{z}) \equiv 
\frac{1}{N}\sum_{a,b\in \bsz_{N}}\, \bar{q}^{\frac{\sn}{2}a^2}
e^{2\pi i \sn a \bar{z}}\,
\prod_{i=1}^{N_M} 
\ch{(\sNS)}{\ell_i,m_i}
   (-\bar{\tau},\bar{z}+a\bar{\tau}+b) \,  \nn
&& \hspace{4cm}\times
 \prod_{j=1}^{N_L} 
\chid^{(\sNS)}(s_j,-K_jn_j+N_jw_j;-\bar{\tau},
-\bar{z}-a\bar{\tau}-b) \nonumber \\
&&\hspace{2cm} \equiv \cG^{(\sNS)}_{I,\wss,\wsw}
(-\bar{\tau},\bar{z})~,
 \nn
&& I= \left((\ell_1,m_1),\ldots, (\ell_{N_M},m_{N_M}), 
n_1,\ldots, n_{N_L}\right)~, ~~~ \s=(s_1,\ldots,s_{N_L})~,~~~
\w=(w_1,\ldots,w_{N_L})~,  \nn
&& \ws=(N_1+2K_1-s_1, \ldots, N_{N_L}+2K_{N_L}-s_{N_L})~,~~~
\ww=(1-w_1,\ldots,1-w_{N_L})~,
\nonumber
\end{eqnarray}
Recall the charge conjugation relations 
\eqn{charge conjugation massless} to derive the last equality.
In the limit $z\rightarrow 0$ we obtain the Witten index 
\begin{eqnarray}
\lim_{z\,\rightarrow\, 0} \, \cG^{(\stR)}_{I,\ss,\sw}(\tau,z) \equiv
\cI_{I,\ss,\sw} ~,
\label{cG WI}
\end{eqnarray}
which can be evaluated by using the formulas \eqn{WI minimal}
and \eqn{Witten index}. 
The elliptic genus is then written as 
\begin{eqnarray}
&&\cZ(\tau,z) = \frac{1}{N} \sum_{I,\ss,\sw}\,\a(\s)\cI_{I,\wss,
\wsw}
\cG^{(\stR)}_{I,\ss,\sw}(\tau,z)~, 
\label{elliptic genus} 
\end{eqnarray}
where we set 
\begin{eqnarray}
&& \a(\s) = \prod_{j=1}^{N_L} \,a(s_j)~, ~~~
a(s_j) = \left\{
\begin{array}{ll}
 1&~~ K_j+1 \leq s_j \leq N_j+K_j-1 \\
 \frac{1}{2} & ~~ s_j=K_j,\, N_j+K_j.
\end{array}
\right. ~~ 
\end{eqnarray}
In the cases of $CY_3$ $(\n=3)$ the elliptic genera
are shown to have a particularly simple form;
\begin{eqnarray}
\cZ(\tau,z) &=& \frac{\chi}{2}\frac{\th_1(\tau,2z)}{\th_1(\tau,z)}~, 
\label{elliptic genus CY3} \\
\chi &= & \frac{1}{N}\sum_{I,\ss,\sw}\, \a(\s) \cI_{I,\ss,\sw} 
\cI_{I,\wss,\wsw} ~. 
\end{eqnarray}

Let us next try to exhibit more explicit forms of 
elliptic genera. 
To this end it is useful to recall the formula of 
elliptic genus of minimal model \cite{Witten-E2}
\begin{eqnarray}
\cZ_{k}(\tau,z)= 
\sum_{\ell=0}^{k}\,\ch{(\stR)}{\ell,\ell+1}(\tau,z) 
=- \sum_{\ell=0}^{k}\,\ch{(\stR)}{\ell,-(\ell+1)}(\tau,z) 
=\frac{\th_1(\tau, \frac{k+1}{k+2}z)}
{\th_1(\tau,\frac{1}{k+2} z)}~.
\label{elliptic genus minimal}
\end{eqnarray}
The corresponding formula for the $L_{N,K}$-sector is written as 
\cite{ES-BH}\footnote
  {Overall sign is opposite to that of \cite{ES-BH}.}
\begin{eqnarray}
\cZ_{N,K}(\tau,z) &\equiv& \sum_{s=K}^{N+K}\,
a(s) \chid^{(\stR)}(s,s-K;\tau,z) \nn 
&\equiv & \left\lb\cK_{2NK}\left(\tau,\frac{z}{N},0 
\right) - \frac{1}{2}\Th{0}{NK}
\left(\tau,\frac{2z}{N}\right) \right\rb \,  
\frac{i\th_1(\tau,z)}{\eta(\tau)^3}~,
\label{cZ N K}
\end{eqnarray}
where $\cK_{\ell}(\tau,\nu,\mu)$ is the level $\ell$ Appell function
\cite{Pol,STT} defined by 
\begin{eqnarray}
\cK_{\ell}(\tau,\nu,\mu) \equiv \sum_{m\in \bsz}\, 
\frac{e^{i\pi m^2 \ell \tau +2\pi i m \ell\nu}}
{1-e^{2\pi i (\nu+\mu+m\tau)}}~.
\label{Appell}
\end{eqnarray}
The following identity is  quite useful;
\begin{eqnarray}
\sum_{s=K}^{N+K-1} \, e^{2\pi i \frac{(s-K)b}{N}}\,
\chid^{(\stR)}(s,s-K+2Ka;\tau,z) &=& 
q^{\frac{K}{N} a^2} y^{\frac{2K}{N}a}\,
\cK_{2NK}\left(\tau,\frac{z+a\tau+b}{N},0\right)\, 
\frac{i\th_1(\tau,z)}{\eta(\tau)^3}~, 
 \nn
&& \hspace{4cm} ~~~ (a,b \in \bz_N)~,
\label{relation chid cK}
\end{eqnarray}
or conversely, 
\begin{eqnarray}
\chid^{(\stR)}(s,s-K+2Ka;\tau,z) &=& \frac{1}{N}
\sum_{b \in \bsz_N} \, e^{-2\pi i \frac{(s-K)b}{N}}\,
q^{\frac{K}{N} a^2} y^{\frac{2K}{N}a}\,
\cK_{2NK}\left(\tau,\frac{z+a\tau+b}{N},0\right)\, 
\frac{i\th_1(\tau,z)}{\eta(\tau)^3}~, \nn
&& \hspace{3.5cm} ~~~ (a \in \bz_N,~~ K\leq s \leq N+K-1)~.
\label{relation chid cK 2}
\end{eqnarray}
One may regard these relations as non-compact analogue of the formula
\eqn{elliptic genus minimal}.
More details on the relation between extended characters and Appell functions 
are discussed in Appendix E.

Our goal is to derive the ``orbifold forms'' of elliptic genera
like those given in \cite{KYY}. 
To this end we have to slightly modify \eqn{cZ N K} 
except for the cases of $N_M=N_L=1$ treated in \cite{ES-BH}, 
so as to correctly reproduce \eqn{elliptic genus}. 
We define 
\begin{eqnarray}
\widehat{\cZ}_{N,K}(\tau,z) &\equiv& \sum_{s=K}^{N+K}\,a(s)
\chid^{(\stR)}(s,s-K-2Kn(s);\tau,z) \nn
&\equiv&  \left\lb \frac{1}{N}\sum_{s=K}^{N+K-1}\,\sum_{b\in \bsz_N}\,
e^{-2\pi i \frac{(s-K)b}{N}} q^{\frac{K}{N}n(s)^2}y^{-\frac{2K}{N}n(s)}\,
\cK_{2NK}\left(\tau, \frac{z-n(s)\tau+b}{N},0\right) 
\right. 
\nn
&& \hspace{2cm} \left. 
-\frac{1}{2} \Th{0}{NK}\left(\tau,\frac{2z}{N}\right)
\right\rb \, \frac{i\th_1(\tau,z)}{\eta(\tau)^3}~,
\label{hat cZ N K}
\end{eqnarray}
where $n(s)$ is defined  uniquely by  the condition
\begin{eqnarray}
&& Kn(s) \equiv s-K ~(\mod\, N)~,~~~ n(s) \in \bz_N~.
\label{ns}
\end{eqnarray}
(This is well-defined for each $s$, since 
we are assuming that $N$ and $K$ are relatively prime.)
In the special case $K=1$, we simply have $n(s)= s-1$. 
The elliptic genus \eqn{elliptic genus} is now rewritten as 
the orbifold form;
\begin{eqnarray}
\cZ(\tau,z) &=& \frac{1}{N} \sum_{a,b\in\bsz_N}\, (-1)^{(N_M+N_L)(a+b)}
q^{\frac{\sn}{2}a^2} y^{\sn a}\,
\prod_{i=1}^{N_M} \cZ_{k_i}(\tau,z+a\tau+b)\, \prod_{j=1}^{N_L}
\widehat{\cZ}_{N_j,K_j}(\tau,z+a\tau+b)~. \nn
&&
\label{elliptic genus orbifold}
\end{eqnarray}

For example, in the case of compactification 
on $ALE(A_{n-1})$ spaces ({\em i.e.} 
$N_M=N_L=1$, $k=n-2$, $N=n$, $K=1$), 
the formula \eqn{elliptic genus orbifold} is reduced to 
\begin{eqnarray}
\cZ_{ALE(A_{n-1})}(\tau,z) &=& \sum_{\ell=0}^{n-2}\, \sum_{r\in \bsz_n}\,
 \ch{(\stR)}{\ell,\ell+1-2r}(\tau,z) \, \chid^{(\stR)}(\ell+2,
\ell+3-2(\ell+2)+2r;\tau,z) \nn
&=& \sum_{\ell=0}^{n-2}\, \sum_{r\in \bsz_n}\, 
\ch{(\stR)}{\ell,-(\ell+1)-2r}
  (\tau,z)\, \chid^{(\stR)}(\ell+2,\ell+1+2r;\tau,z)~,
\label{elliptic genus ALE}
\end{eqnarray}
which reproduces the one given in \cite{ES-BH}.

~

\subsection{Massless Closed String Spectra}

~

It is an important task to analyze the massless closed string spectrum. 
We can solve this problem in a similar manner as in 
the compact Gepner models, since we have already constructed 
the conformal blocks in closed string sector.  
Massless states correspond to the (anti-) chiral primary states 
of conformal weights $h=\tilde{h}=1/2$.
As we discuss below, 
basic aspects of massless spectra in the non-compact models 
are summarized as follows:  
\begin{itemize}
 \item In the cases of $\hat{c}\neq 2$, there exists 
at most one chiral primary of the $(a,c)$ (or $(c,a)$)-type
with $h=\tilde{h}=1/2$ in each spectral flow orbit \eqn{cG},
and none of the $(c,c)$ (or $(a,a)$)-type exist \footnote
   {Of course, in mirror models where $L_{N_j,K_j}$-sectors
     are realized by $\cN=2$ Liouville theories, 
     the situation is reversed
     : no chiral primaries of $(c,a)$ and $(a,c)$-types exist, while  
     $(c,c)$ and $(a,a)$-types are possible.}. 
 \item In the cases of $\hat{c}=2$, we have at most a quartet
of the $(c,c)$, $(a,a)$, $(c,a)$ and $(a,c)$-type primaries 
with $h=\tilde{h}=1/2$ in each spectral flow orbit. 
\end{itemize}
This fact implies that at our non-compact Gepner points for 
$CY_3$ or $CY_4$ moduli space there exist  
only the deformations of K\"{a}hler structure but not the deformations of
complex structure. On the other hand, 
in the case of $K3$ surfaces $\hat{c}=2$ 
the superconformal symmetry is extended to $\cN=4$
and the above massless states compose 
the spin (1/2,1/2) representation of $SU(2)_L \times SU(2)_R$ 
of the $\cN=4$ SCA. From the space-time point of view 
the quartet corresponds to a scalar (tensor)-multiplet of $(1,1)$ ($(2,0)$) 
SUSY in 6 dimensions.

Let us consider the $(a,c)$-type massless states: 
the analysis for the $(c,a)$-type is parallel. 
We start with working on the left-moving sectors.
The anti-chiral states are described by the conditions;
\begin{list}{}
 \item $M_{k_i}$-sector :
\begin{eqnarray}
m_i= -\ell_i~, ~~(0\leq \ell_i \leq k_i)~,
~~~ \mbox{$U(1)$-charge    }~ Q_i= -\frac{\ell_i}{k_i+2}~,
\label{anti-chiral min}\end{eqnarray}
 \item $L_{N_j,K_j}$-sector :
\begin{eqnarray}
&& s_j=K_jn_j+N_jw_j+2K_j~,~~(K_j \leq s_j \leq N_j+K_j)~,
~~~ 
m_j=K_jn_j+N_jw_j~, \nn
&& \mbox{$U(1)$-charge    }~ Q_j= \frac{K_jn_j+N_jw_j-N_j}{N_j}~,
\label{anti-chiral coset}\end{eqnarray}
\end{list}{} 
and we must impose 
\begin{eqnarray}
-\sum_i \frac{\ell_i}{k_i+2} + \sum_j \frac{K_jn_j+N_jw_j-N_j}{N_j} =-1~.
\label{cond anti-chiral}
\end{eqnarray}
We now recall the left-right pairing of states in our construction is given as 
 (see \eqn{cG});
 \begin{equation}
 \begin{array}{ccc}
 \mbox{left-moving} & \null & \mbox{right-moving} \\
 (\ell_i,m_i) &\Longleftrightarrow &(\tilde{\ell}_i=\ell_i,\tilde{m}_i=m_i)\\
 (s_j,m_j=K_jn_j+N_jw_j) & \Longleftrightarrow &
  (\tilde{s}_j=N_j+2K_j-s_j,\tilde{m}_j=K_jn_j-N_jw_j+N_j)
 \label{left-right pairing}
 \end{array}
\end{equation}
Thus before applying the spectral flow the right-moving state corresponding to 
(\ref{anti-chiral min}), (\ref{anti-chiral coset})
has the following quantum numbers
\begin{eqnarray}
(\tilde{\ell}_i, \tilde{m}_i) 
=(\ell_i, -\ell_i) ~, ~~~ (\tilde{s}_j, \tilde{m}_j) 
= (-K_jn_j-N_jw_j+N_j, K_jn_j-N_jw_j+N_j)~. 
\label{right mover 1}
\end{eqnarray}
They have the total $U(1)$-charge 
\begin{eqnarray}
-\sum_i \frac{\ell_i}{k_i+2} + \sum_j \frac{K_jn_j-N_jw_j+N_j}{N_j}~,
\end{eqnarray}
which is integer because of the constraint \eqn{cond anti-chiral}.

Now, we look for chiral or anti-chiral 
primaries with $\tilde{h}=1/2$ in the orbit of spectral flow starting from 
the states \eqn{right mover 1}. 
The flow $\bar{z}\,\rightarrow\,\bar{z}+\bar{\tau}$
acts on the quantum numbers as 
\begin{eqnarray}
\tm_i~\longrightarrow~\tm_i-2~, ~~~ \tm_j~\longrightarrow~\tm_j+2K_j~,
\end{eqnarray}
and it has periodicities $k_i+2$ and $N_j$ respectively in the $M_{k_i}$ and 
$L_{N_j,K_j}$-sectors.  
We find that only the orbits satisfying 
the condition
\begin{eqnarray}
{}^{\exists} r \in \bz_N~, ~ \mbox{s.t.}~ 
\left\{
\begin{array}{l}
 \ell_i \equiv r~(\mod\, k_i+2)~, ~~ {}^{\forall}i \\
 n_j \equiv r ~ (\mod\, N_j)~,~~ {}^{\forall}j
\end{array}
\right. 
\label{r constraint}
\end{eqnarray}
contains a (unique) chiral state 
with the total $U(1)$-charge $Q_{\msc{tot}}=1$;
\begin{eqnarray}
(\tilde{\ell}_i,\tilde{m}_i)=(\ell_i,\ell_i)~,~~~ 
(\tilde{s}_j,\tilde{m}_j)= (-K_j n_j +N_j(1-w_j),-K_j n_j +N_j(1-w_j))~,  
~({}^{\forall}i,{}^{\forall}j).
\label{chiral state}
\end{eqnarray}
It  also contains 
a (unique) anti-chiral state
with $Q_{\msc{tot}}=1-\hat{c}$;
\begin{eqnarray}
&&(\tilde{\ell}_i,\tilde{m}_i)=(\ell_i,\ell_i+2) \cong 
(k_i-\ell_i, \ell_i-k_i)~,~~~ \nn
&&(\tilde{s}_j,\tilde{m}_j)
= (-K_j n_j +N_j(1-w_j),-K_j (n_j+2) +N_j(1-w_j))~,  
~({}^{\forall}i,{}^{\forall}j)~.
\label{anti-chiral state}
\end{eqnarray}
In fact, \eqn{chiral state}, \eqn{anti-chiral state} 
are generated by the spectral flows 
$\bar{z}\,\rightarrow\,\bar{z}-r \bar{\tau}$,
$\bar{z}\,\rightarrow\,\bar{z}-(r+1)\bar{\tau}$
from \eqn{right mover 1} respectively if \eqn{r constraint} holds. 
Note that in the cases of $\hat{c}=3,4$ 
only \eqn{chiral state} yields a massless state, while both of 
\eqn{chiral state}, \eqn{anti-chiral state} become 
massless states at $\hat{c}=2$.

Therefore, the spectrum of massless closed string states 
is given by the solutions $(\ell_i, n_j, w_j) $ ($0\leq \ell_i \leq
k_i$,  $n_j\in \bz_{N_j}$,  $w_j \in \bz_{2K_j}$)
of the constraints
\begin{eqnarray}
&& \left(\sum_i \frac{\ell_i}{k_i+2}-\sum_j\frac{K_j n_j}{N_j}\right)
= 1+ \sum_j(w_j-1) ~,
\label{charge constraint}
\\
&& -K_j \leq K_jn_j+N_jw_j \leq N_j-K_j~,
\label{s constraint}
\end{eqnarray}
as well as \eqn{r constraint}. 
The second condition \eqn{s constraint} follows from the constraint
$K_j \leq s_j \leq N_j+K_j$.
Obviously, the counting of $(c,a)$-type chiral states can be 
carried out in the same way, yielding the equal number 
of massless states. We have thus shown the characteristic feature of 
massless states announced before. 

We now present concrete examples that have clear geometrical 
interpretations.

~


\noindent
{\bf 1. Cases of one Liouville field $N_L=1$ : }

We first present examples
with $N_M=N_L=1$.
The condition \eqn{r constraint} simply yields $\ell_1=n_1(\equiv \ell)$
in these cases.

~

\begin{description}
 \item[1.1. ALE($A_{n-1}$) : $M_{n-2} \otimes L_{n,1}$] 

~

In this case
the constraint \eqn{charge constraint} simply gives $w_1=0$, and 
\eqn{s constraint} is equivalent with 
\begin{eqnarray}
 -1 \leq \ell  \leq n-1~.
\end{eqnarray}
We thus conclude that each (anti-)chiral primary state 
in the range $0\leq \ell \leq n-2$ in $M_{n-2}$
can be paired up to massless states 
of the types of $(c,c)$, $(a,a)$, $(c,a)$ and 
$(a,c)$-types of $M_{n-2} \otimes L_{n,1}$ theory.

\item[1.2. $CY_4$ ($A_{n-1}$) : $M_{n-2}\otimes L_{n,n+1}$]

~

The condition \eqn{charge constraint} is solved as
\begin{eqnarray}
\ell   = -w_1~, ~~~ w_1 \in \bz_{2(n+1)}. 
\end{eqnarray}
\eqn{s constraint} then gives 
\begin{eqnarray}
-(n+1) \leq \ell \leq -1~,
\end{eqnarray}
which has no solution in the range $0\leq \ell \leq n-2$.
Therefore, we have no massless states in this case.

\item[1.3. $CY_3$ ($A_{n-1}$) : $M_{n-2}\otimes L'_{2n,n+2}$]

~

This case is non-trivial. 
We have $N_1=2n$, $K_1=n+2$, which are not necessarily relatively prime.
As addressed before, we must 
allow the half-integral winding numbers $w_1$, and impose the constraint
$n_1+2w_1 \in 2\bz$. 

The constraint \eqn{charge constraint}
now leads to 
\begin{eqnarray}
 \ell = -2w_1~, ~~~ 
\left\{
\begin{array}{ll}
w_1\in \bz_{2(n+1)}~ & ~~ \ell~:~ \mbox{even} \\
w_1\in \frac{1}{2} + \bz_{2(n+1)}~ & ~~ \ell~:~\mbox{odd}
\end{array}
\right.
\end{eqnarray}
and \eqn{s constraint} gives 
\begin{eqnarray}
-(n+2) \leq 2\ell \leq n-2~.
\end{eqnarray}
We thus find that the (anti-)chiral states in $M_{n-2}$
with 
\begin{eqnarray}
\ell = 0, 1, \ldots, \left\lb \frac{n-2}{2} \right\rb
\label{massless singular CY3}
\end{eqnarray}
produce massless states of the $(c,a)$ and $(a,c)$-types.

\end{description}

~

As already mentioned in \cite{ES-BH}, these aspects of massless states
in the above three examples are consistent with the spectra of  
{\em normalizable\/} chiral operators describing the moduli of vacua 
discussed in \cite{GVW,GKP,Pelc}. Especially, 
the third example correctly reproduces the spectrum of
scaling operators in the $\cN=2$ $SCFT_4$ of Argyres-Douglas points 
\cite{AD}. 
We also point out that the massless 
spectra here are consistent with the ones deduced from the ``LSZ poles"
in correlation functions presented in \cite{AGK}.

~

\noindent
{\bf 2. Cases of two Liouville fields $N_L=2$ : }

\begin{description}
 \item[2.1. $\hat{c}=3$, $M_{n-2}\otimes L_{2n,1} \otimes L_{2n,1}$ :  ]


~

The criticality condition is satisfied as 
\begin{eqnarray}
\hat{c} = \frac{n-2}{n} + \left(1+\frac{1}{n}\right)+
\left(1+\frac{1}{n}\right) = 3~.
\end{eqnarray}
This type of superconformal system has been first studied in \cite{Lerche}
and proposed to be the superstring vacuum corresponding to the 
Seiberg-Witten theory with $SU(n)$ gauge group without matter  
in the low energy regime. Geometrically the theory is supposed to describe 
space-time of 
the $ALE(A_{n-1})$-fibration over $\bc P^1$, or the $n$ NS5-branes 
wrapped around $\bc P^1$ in the T-dual picture.

We have two possibilities of satisfying  \eqn{r constraint};
\begin{list}{}
 \item (i) $\ell_1=n_1=n_2 \equiv r$, $0\leq r \leq n-2$,
 \item (ii) $\ell_1+n=n_1=n_2 \equiv r$, $n\leq r \leq 2n-2$.
\end{list}
In the case (i), \eqn{charge constraint} gives $w_1+w_2=1$,
and \eqn{s constraint} leads to 
\begin{eqnarray}
-1 \leq r +2n w_i \leq 2n-1~,~~(i=1,2)~.
\label{constraint ex 2}
\end{eqnarray}
There are no solutions to these constraints.

In the case (ii), \eqn{charge constraint} gives $w_1+w_2=0$.
If (and only if) we set $w_1=w_2=0$,  \eqn{constraint ex 2} 
are satisfied for arbitrary $n\leq r \leq 2n-2$. 
We thus find that 
\begin{eqnarray}
  \ell_1=0,1,\ldots, n-2~,~~
 n_1=n_2=\ell_1+n~,~~ w_1=w_2=0~,~ 
\Longleftrightarrow~\mbox{massless states}~.
\end{eqnarray}
The $(c,a)$-type chiral fields also gives the equal number of 
massless states.  These are identified as the 
moduli $u_2,\ldots, u_{n}$ in the $SU(n)$ SW theory. 
Especially, the marginal deformation for $\ell=0$  
corresponds to the size of base $\bc P^1$ as suggested in \cite{HK2}.

\item[2.2.   $\hat{c}=3$, $M_{n-2} \otimes L_{n\mu,K_1} \otimes 
L_{n\mu,K_2}$, $K_1+K_2=\mu$, $\mbox{G.C.D}\{K_i\}=1$ : ]

~

This is a natural generalization of the example {\bf 2.1}. 
The criticality condition is satisfied as 
\begin{eqnarray}
\hat{c} \equiv \left(1-\frac{2}{n}\right) 
+ \left(1+\frac{2K_1}{N_1}\right)
+ \left(1+\frac{2K_2}{N_2}\right) = 
3 + \frac{2}{N}\left(-\mu  + K_1+K_2\right) =3~.
\end{eqnarray}
In this case, using the relation $K_1+K_2=\mu$, we again obtain
\begin{eqnarray}
  \ell_1=0,1,\ldots, n-2~,~~
  n_1=n_2=\ell_1+n~,~~ w_1=w_2=0~,~ 
\Longleftrightarrow~\mbox{massless states}~.
\end{eqnarray}
This type of string vacua are identified as the non-compact $CY_3$
with the structure of $ALE(A_{n-1})$-fibration over 
the weighted projective space $W\bc P^1 \left\lb K_1,K_2\right\rb$.

\item[2.3. $\hat{c}=4$, $M_{n-2} \otimes L'_{4n,n+2} \otimes L'_{4n,n+2}$ : ]

~

We next consider a more subtle example. 
The criticality condition is satisfied as 
\begin{eqnarray}
\hat{c} = \frac{n-2}{n} + \left(1+\frac{n+2}{2n}\right)+
\left(1+\frac{n+2}{2n}\right) = 4~.
\end{eqnarray}
Similarly to the case of $CY_3 (A_{n-1})$, we have to allow  
$\dsp w_j \in \frac{1}{4}\bz$ and assume \eqn{m CY3 fiber} 
in each of $L'_{4n,n+2}$-sector
in order to obtain the expected spectrum. 
As in the example 2.1, massless states are possible only for 
\begin{eqnarray}
\ell_1+n = n_1=n_2 = r~, ~~~ 0\leq r \leq 2n-2~, ~~~ r+4w_i=0~,
\end{eqnarray}
and \eqn{s constraint} gives us 
\begin{eqnarray}
\ell = 0, 1, \ldots, \left\lb \frac{n-2}{2} \right\rb
~ \Longleftrightarrow~ \mbox{massless states}.
\end{eqnarray}
This spectrum is the same as  $CY_3$ ($A_{n-1}$).
We propose that this model is identified as 
$CY_3 (A_{n-1})$-fibration on $\bc P^1$.

The generalizations similar to the example {\bf 2.2} 
are straightforward; $M_{n-2} \otimes L'_{2n\mu, K_1(n+2)} \otimes 
L'_{2n\mu, K_2(n+2)}$, $K_1+K_2=\mu$, $\mbox{G.C.D}\{K_i\}=1$. 
It is expected that it describes the $CY_3(A_{n-1})$-fibration over 
$W\bc P^1 \left\lb K_1,K_2\right\rb$ and we again obtain 
the same massless spectrum \eqn{massless singular CY3}.

\end{description}

~

\noindent
{\bf 3.  Cases of three Liouville fields $N_L=3$ : }

The $\hat{c}=4$ vacua are the only possibility for these cases. 

\begin{description}
 \item[3.1. $M_{n-2} \otimes L_{3n,1}\otimes L_{3n,1} \otimes L_{3n,1}$ : ]

~

The criticality condition is satisfied as 
\begin{eqnarray}
\hat{c}= \frac{n-2}{n} + 3\times \left(1+ \frac{2}{3n}\right) =4~.
\end{eqnarray}
The massless states are possible only for 
\begin{eqnarray}
\ell_1+2n = n_1=n_2=n_3 = r~, ~~~ 2n \leq r \leq 3n-2~, ~~~ w_1=w_2=w_3=0~.
\end{eqnarray}
\eqn{s constraint} gives us 
\begin{eqnarray}
\ell_1= 0, 1, \ldots, n-2~ \Longleftrightarrow~ \mbox{massless states}~,
\label{massless ALE fiber P2}
\end{eqnarray}
which is the same spectrum as that of $ALE(A_{n-1})$.
However, we can only have the $(a,c)$ and $(c,a)$-type massless chiral 
states contrary to the ALE case. 
This model is identified as the $ALE(A_{n-1})$-fibration over $\bc P^2$
\cite{HK2}.


The generalization similar to the example {\bf 2.2} is also easy;
$M_{n-2} \otimes L_{n\mu,K_1} \otimes L_{n\mu, K_2} \otimes L_{n\mu,K_3}$,
$K_1+K_2+K_3=\mu$, $\mbox{G.C.D}\{K_i\}=1$.
This model is identified as the $ALE(A_{n-1})$-fibration over 
$W \bc P^2 \left\lb K_1,K_2,K_3\right\rb$ and 
we again obtain the same massless spectrum \eqn{massless ALE fiber P2}.

\end{description}

~


\subsection{Notes on Geometrical Interpretations}

~

Let us here clarify our geometrical interpretation of
the string vacua considered above 
as non-compact Calabi-Yau spaces. 
We first recall the familiar CY/LG correspondence:
\begin{eqnarray}
X_1^{r_1}+ \cdots + X_{n+2}^{r_{n+2}} =0~,~~~ 
\mbox{in}~ W\bc P_{n+1} \left\lb \frac{1}{r_1}, \ldots, \frac{1}{r_{n+2}} 
\right\rb~, 
\label{CY general}
\end{eqnarray}
defines a Calabi-Yau $n$-fold if $\dsp \sum_{i=1}^{n+2} \frac{1}{r_i} =1$,
and is equivalent to the LG orbifold 
defined by the superpotential 
$W(\X_i) \equiv \X_1^{r_1} + \cdots + \X_{n+2}^{r_{n+2}}$,
where $\X_i$ denotes the chiral superfields.  
If some of $r_i$ are negative, the corresponding Calabi-Yau space
becomes non-compact. 
In fact, such LG interpretation was 
the starting point for the CFT descriptions of 
non-compact Calabi-Yau spaces \cite{GV,OV,GKP,Lerche}, 
and was further refined from the viewpoints of mirror symmetry in 
\cite{HK2}.

As an illustration, we consider the example {\bf 2.1} : 
$\hat{c}=3$, $M_{n-2}\otimes L_{2n,1} \otimes L_{2n,1}$.
The corresponding LG model is given as \cite{Lerche}
\begin{eqnarray}
W= \X^n + \Y_1^{-2n} + \Y_2^{-2n}~,
\label{LG Lerche}
\end{eqnarray}
which describes the non-compact CY space
\begin{eqnarray}
X^n + Y_1^{-2n} + Y_2^{-2n} + w_1^2 + w_2^2 =0~,~~~ \mbox{in}~ 
W \bc P^4 \left\lb 2, -1, -1, n, n \right\rb~.
\label{CY Lerche}
\end{eqnarray}
This formula has structure of the $ALE(A_{n-1})$-fibration over $\bc P^1$.
Following  \cite{GKP}, one may 
rewrite \eqn{LG Lerche} as the Liouville form \footnote
   {Here we absorbed the cosmological constants $\mu_1$, $\mu_2$ 
    by shifting zero-modes of $\X_1$, $\X_2$.}; 
\begin{eqnarray}
W= \X^n + e^{-\frac{1}{\cQ} \sX_1} + e^{-\frac{1}{\cQ} \sX_2}~,
\label{LG Lerche 2}
\end{eqnarray}
where we set $\cQ=\sqrt{1/n}$. 
This realization 
amounts to expressing the 
$L_{2n,1}$-sectors as the $\cN=2$ Liouville theories, and the linear
dilaton is given by
\begin{eqnarray}
\Phi = -\frac{\cQ}{2} \Re\,\left(\X_1+\X_2\right)~.
\end{eqnarray} 
Equivalently, one may rewrite it as
\begin{eqnarray}
W= \X^n + e^{-n \sZ} (e^{\sY}+ e^{-\sY}) ~, 
\label{LG Lerche 3}
\end{eqnarray}
where we set $\X_1=n\cQ \Z + \cQ \Y$, $\X_2= n\cQ \Z -\cQ \Y$.
In this parameterization the linear dilaton is 
along the $\Z$-direction;
\begin{eqnarray}
\Phi = -\Re\,\Z~.
\end{eqnarray}
One can directly recover the geometry of $ALE(A_{n-1})$-fibration over 
$\bc P^1$;
\begin{eqnarray}
e^{Y}+ e^{-Y} + X^n+w_1^2+w_2^2=0~,
\end{eqnarray}
by integrating out the chiral superfield $\Z$ (with the rescaling 
$\X\, \rightarrow\, e^{-\sZ}\X$)   \cite{HK2}.

Similarly, the example {\bf 3.1}:
$\hat{c}=4$, $M_{n-2}\otimes L_{3n,1} \otimes L_{3n,1} \otimes L_{3n,1}$
can be identified as the LG theory with 
\begin{eqnarray}
W= \X^n + e^{-n \sZ} (e^{\sY_1}+ e^{\sY_2}+e^{-\sY_1-\sY_2}) ~, 
\label{LG P2 fibration}
\end{eqnarray}
where we have again the linear dilaton $\Phi = - \Re\, \Z$. 
This model is shown to describe the $ALE(A_{n-1})$ fibration 
over $\bc P^2$ \cite{HK2}. 

Other examples are also identified in a similar
manner, leading to the geometrical interpretations mentioned above. 
For instance, the model {\bf 2.2} : 
$\hat{c}=3$, $M_{n-2}\otimes L_{n\mu,K_1}\otimes L_{n\mu,K_2}$ 
($K_1+K_2=\mu$) is identified with the LG theory with
\begin{eqnarray}
W= \X^n + e^{-n \sZ} (e^{\sY / K_1}+ e^{-\sY / K_2}) ~, ~~~ 
\Phi = -\Re\,\Z~.
\label{LG WP1}
\end{eqnarray}
It corresponds to the non-compact Calabi-Yau geometry
\begin{eqnarray}
e^{Y / K_1}+ e^{-Y /K_2} + X^n+w_1^2+w_2^2=0~,
\end{eqnarray}
which has the structure of $ALE(A_{n-1})$-fibration over 
$W\bc P^1 \left\lb K_1,K_2\right\rb$.

~


\section{D-branes in Non-compact Models}

~

\subsection{Cardy States for Compact BPS D-branes}

~

We next study the open string sectors in the non-compact Gepner models. 
It is well-known  that the $\cN=2$ superconformal symmetry allows 
two types of boundary conditions \cite{OOY} \footnote
  {A slightly different convention is often used 
in literature; A-type brane, for instance, is defined by
$(G^{\pm}_r-i \eta \tG^{\mp}_{-r})\ket{B;\eta}=0~,~ (\eta =\pm 1)~.$
The relation to our convention is given by 
$ \ket{B;+1}= \ket{B}~,~ \ket{B;-1}= (-1)^{F_R} \ket{B}~.$
};
\begin{eqnarray}
&&  \mbox{\bf A-type}~~~ :~
(J_n-\tJ_{-n})\ket{B}=0 ~,~~~(G^{\pm}_r-i\tG^{\mp}_{-r})\ket{B}=0~,
\label{A-type} \\
&& \mbox{\bf B-type}~~~ :~
 (J_n+\tJ_{-n})\ket{B}=0 ~,~~~(G^{\pm}_r-i\tG^{\pm}_{-r})\ket{B}=0~,
\label{B-type}
\end{eqnarray}
which are compatible with the $\cN=1$ superconformal symmetry
\begin{eqnarray}
(L_n-\tL_{-n})\ket{B}=0~,~~~ (G_r-i\tG_{-r})\ket{B}=0~,
\label{N=1 gluing}
\end{eqnarray}
where $G=G^++G^-$ is the $\cN=1$ supercurrent.
In the following we shall concentrate on the BPS D-branes 
with compact world-volumes,
which play the fundamental role  in the non-compact 
Calabi-Yau manifolds. 
Compact branes in $L_{N_j,K_j}$-sectors are described by the `class 1' 
Cardy states in the classification of \cite{ES-L}, namely, 
the ones associated to the extended graviton representations. 
It can be shown  that the consistency with the left-right pairing of the 
closed string spectrum (\ref{left-right pairing})
allows 
{\em only\/} the B-type boundary condition  
for these compact branes \cite{RibS,IPT2,FNP}.

We will begin our analysis 
by summarizing characteristic aspects of boundary states 
in each sector\footnote
  {Here we use notations slightly different from \cite{ES-L}.}. 
We assume in the following that 
all the Ishibashi states satisfy the B-type boundary condition.

~

\noindent
{\bf 1. minimal sector $M_k$ : } (see {\em e.g.} \cite{RS})

Let $\dket{\ell,m}^{(\sNS)}$
($\dket{\ell,m}^{(\sR)}$) be 
the Ishibashi states in the NS (R) sector
characterized by the orthogonality condition ($\sigma = \NS,\, \R$)
\begin{eqnarray}
&& \hspace{-1.3cm} {}^{(\sigma)}\dbra{\ell,m} e^{-\pi T H^{(c)}}
e^{2 \pi i z J_0}
 \dket{\ell',m'}^{(\sigma')} 
= \ep_{\sigma}\delta_{\sigma,\sigma'}
\left(\delta_{\ell,\ell'}\delta_{m,m'}
+\delta_{\ell,k-\ell'}\delta_{m,m'+k+2}
\right) \, \ch{(\sigma)}{\ell,m}(iT,z), 
\end{eqnarray}
where $H^{(c)}=L_0+\tilde{L}_0-{c\over 12}$ 
is the closed string Hamiltonian 
and $\ch{(\sNS)}{\ell,m}(\tau,z)$ ($\ch{(\sR)}{\ell,m}(\tau,z)$) 
denotes the NS (R) character of the $\cN=2$
minimal model for the primary field with
$\dsp h=\frac{\ell(\ell+2)-m^2}{4(k+2)}$, $\dsp Q=\frac{m}{k+2}$~
($\dsp h=\frac{\ell(\ell+2)-m^2}{4(k+2)}+\frac{1}{8}$, 
$\dsp Q= \frac{m}{k+2}\pm \frac{1}{2}$). (See Appendix B.)
We have introduced an extra phase factor $\ep_{\sigma}=+1,-1$ for 
$\sigma=\NS, \R$ respectively for convenience in imposing 
the GSO projection for supersymmetric D-branes. 
We also set $\dket{\ell,m}^{(\sNS)}=0$
($\dket{\ell,m}^{(\sR)}=0$), if $\ell+m \in 2\bz+1$ ($\ell+m \in 2\bz$).
The  Cardy states  are expressed as follows
($\sigma=\NS, ~ \mbox{or} ~ \R~$, and we set $L+M \in 2\bz$);
\begin{eqnarray}
&&\ket{L,M}^{(\sigma)} = \sum_{\ell=0}^k\,\sum_{m\in \bsz_{2(k+2)}}\,
C_{L,M}(\ell,m)\dket{\ell,m}^{(\sigma)}~, ~~~
 C_{L,M}(\ell,m) = \frac{S^{L,M}_{\ell,m}}{\sqrt{S^{0,0}_{\ell,m}}}~,
\label{minimal Cardy states}
\end{eqnarray}
where $S^{\ell,m}_{\ell',m'}$ is the modular coefficients of 
$\ch{(\sNS)}{\ell,m}(\tau,z)$ \eqn{minimal S}.

~

\noindent
{\bf 2. $SL(2;\br)/U(1)$-sector $L_{N,K}$ : }

The relevant formulas of boundary states in this sector 
are given in \cite{ES-L}. 
The (B-type) Ishibashi states corresponding 
to continuous and discrete representations 
are characterized by the relations
\begin{eqnarray}
&& \hspace{-1cm}
{}^{(\sigma)}\dbrac{p,m} e^{-\pi T H^{(c)}} e^{ 2\pi i  z J_0}
\dketc{p',m'}^{(\sigma')} = \ep_{\sigma}
\delta_{\sigma,\sigma'} \delta_{m,m'}^{(2NK)} 
\delta(p-p')\, \chic^{(\sigma)}(p,m;iT,z) ~, \nn
&& \hspace{8cm} (p, p'>0)~, 
\label{Ishibashi cont}\\
&& \hspace{-1cm}
{}^{(\sigma)}\dbrad{s,m} e^{-\pi T H^{(c)}} e^{ 2 \pi i  z J_0}
\dketd{s',m'}^{(\sigma')} = \ep_{\sigma} 
\delta_{\sigma,\sigma'} \delta_{m,m'}^{(2NK)} 
\delta_{s,s'}\, \chid^{(\sigma)}(s,m;iT,z) ~,
\label{Ishibashi dis}
\end{eqnarray}
where the range of $s$ is $K+1 \leq s \leq N+K-1$. 
We need not introduce Ishibashi states at the boundary 
$s=K,\,N+K$ as is discussed in \cite{ES-L}.
We here set 
\begin{eqnarray}
&& \hspace{-1cm}
\dketd{s,m}^{(\sNS)} = 0~,~\mbox{unless } s-m \in 2K \bz~, ~~~
\dketd{s,m}^{(\sR)} = 0~,~\mbox{unless } s-m \in K(2\bz+1)~. 
\end{eqnarray}

The Cardy states necessary for our analysis are the class 1 states
given in \cite{ES-L} ($R\in \bz_N$)\footnote
  {Recall that our closed string Hilbert space for the 
  $L_{N,K}$-piece includes the twisted sectors 
  generated by spectral flows unlike the `cigar CFT' 
  given in \cite{RibS,IPT2,FNP}. 
  Note that the D0-brane of cigar CFT corresponds to the boundary state
$
 \sum_{R\in \bsz_{N}} 
 \ket{R}^{(\sigma)}
$
in our notation, 
where the summation over $R \in \bz_{N}$ 
eliminates the twisted sectors. 
}
;
\begin{eqnarray}
&& \hspace{-1cm}
\ket{R}^{(\sigma)} = \sum_{s=K+1}^{N+K-1}\, \sum_{m\in \bsz_{2NK}}\,
C^{(\sigma)}_R(s,m) \dketd{s,m}^{(\sigma)} + 
\sum_{m\in \bsz_{2NK}}\, \int_0^{\infty}dp\, \Psi^{(\sigma)}_R(p,m) 
\dketc{p,m}^{(\sigma)}~, 
\label{Cardy L}\\
&& \hspace{-1cm}
C_R^{(\sNS)}(s,m) = C_R^{(\sR)}(s,m) = \left(\frac{2}{N}\right)^{1/2} 
e^{-2\pi i \frac{Rm}{N}} \sqrt{\sin \left(\frac{\pi (s-K)}{N}\right)}~, 
\label{C R} \\
&& \hspace{-1cm}
\Psi_R^{(\sigma)}(p,m) = \left(\frac{2^3 K}{N^3}\right)^{1/4} 
e^{-2\pi i \frac{Rm}{N}}\, 
\frac{\Gamma\left(\frac{1}{2}+\frac{m-\nu(\sigma)K}{2K}
+i\sqrt{\frac{N}{2K}}p\right)
\Gamma\left(\frac{1}{2}-\frac{m-\nu(\sigma)K}{2K}+i\sqrt{\frac{N}{2K}}p\right)}
{\Gamma\left(i\sqrt{\frac{2K}{N}}p\right) \Gamma(1+i\sqrt{\frac{2N}{K}}p)}~,
\label{Psi R}
\end{eqnarray}
where the symbol $\nu(\sigma)$ means $\nu(\NS)=0$, $\nu(\R)=1$.
Cylinder amplitudes of the class 1 Cardy states produce characters of identity representations;
\begin{eqnarray}
&&  e^{\pi \hat{c} \frac{z^2}{T}}\cdot {}^{(\sNS)}\bra{R} e^{-\pi T H^{(c)}}
e^{ 2 \pi i z J_0} \ket{R'}^{(\sNS)}
= \chi_0^{(\sNS)}(2K(R'-R); it , z') ~, \\
&&  e^{\pi \hat{c} \frac{z^2}{T}}\cdot {}^{(\sR)}\bra{R} e^{-\pi T H^{(c)}}
e^{2 \pi i  z J_0} \ket{R'}^{(\sR)}
= \chi_0^{(\stNS)}(2K(R'-R); it , z') ~, \\
&& \hspace{3cm} (T \equiv 1/t~,~~~ z'= -it z)~.
\nonumber
\end{eqnarray}

~


The desired Cardy states for our non-compact Gepner model
\eqn{nc gepner} should be constructed as 
\begin{eqnarray}
&& \ket{B; \{L_i, M_i\}, \{R_i\}; \pm} = 
\ket{B; \{L_i, M_i\}, \{R_i\}}^{(\sNS)} \pm 
\ket{B; \{L_i, M_i\}, \{R_i\}}^{(\sR)}~, \nn
&& \ket{B; \{L_i,M_i\}, \{R_i\}}^{(\sigma)} = \cN\, 
P_{\msc{closed}}\, \left\lb \prod_i \ket{L_i,M_i}^{(\sigma)}
 \otimes \prod_j \ket{R_j}^{(\sigma)}\right\rb~.
\label{Cardy ncg}
\end{eqnarray}
In the first equation $\pm$ refers to branes and anti-branes, respectively. 
$\cN$ is an overall normalization  constant 
determined by the Cardy condition (its explicit value is not
important for our analysis), and 
$P_{\msc{closed}}$ means the projection to the closed string Hilbert 
space determined in our previous analysis. Namely, $P_{\msc{closed}}$
imposes the following two constraints on the Ishibashi states
$\dket{\ell_i,m_i}^{(\sigma)}$, $\dketc{p_j,K_jn_j+N_jw_j}^{(\sigma)}$, 
and $\dketd{s_j,K_jn_j+N_jw_j}^{(\sigma)}$;
\begin{itemize}
 \item Integrality of the total $U(1)$-charge in the $\NS$-sector 
;
\begin{eqnarray}
\sum_{i}\frac{m_i}{k_i+2} + \sum_j\frac{K_jn_j}{N_j} \in \bz~.
\label{U(1) integrality}
\end{eqnarray}
 \item The consistency with the B-type boundary condition,
which gives essentially the same condition as \eqn{r constraint};
\begin{eqnarray}
{}^{\exists}r \in \bz_N~~ \mbox{s.t.}~
  m_i \equiv -r ~(\mod\, k_i+2)~, ~~~ n_j \equiv r ~ (\mod\, N_j)~,~~
 ({}^{\forall}i,{}^{\forall}j)~.
\label{r constraint 2}
\end{eqnarray}
\end{itemize}
(\ref{r constraint 2}) is derived as follows;
due to the left-right pairing in our construction (\ref{left-right pairing})
a typical state in the closed string spectrum has the form
\begin{equation}
\begin{array}{l}
|\ell_i,m_i\rangle\otimes |\tilde{\ell_i}=\ell_i,\tilde{m_i}=m_i\rangle: 
\hskip2cm  \mbox{in the  } M_{k _i}\mbox{-sector},\\
|s_j,m_j=K_jn_j+N_jw_j\rangle\otimes|\tilde{s_j}
=N_j+2K_j-s_j,\tilde{m_j}=K_jn_j-N_jw_j+N_j\rangle: \\
\hskip9cm  \mbox{in the  }L_{N_j,K_j} \mbox{-sector}
\end{array}
\label{initial state}
\end{equation}
Corresponding B-type boundary state has the quantum numbers as
 \begin{equation}
\begin{array}{l}
|\ell_i,m_i\rangle\otimes |\tilde{\ell_i}=\ell_i,\tilde{m_i}=-m_i\rangle, \\
|s_j,m_j=K_jn_j+N_jw_j\rangle\otimes|\tilde{s_j}=N_j+2K_j-s_j,\tilde{m_j}=-K_jn_j-N_jw_j+N_j\rangle.
\label{B-type state}\end{array}
\end{equation}
(\ref{B-type state}) can be obtained from (\ref{initial state}) by spectral flow in the right moving sector if
(\ref{r constraint 2}) is obeyed. 
Analysis of massive Ishibashi states is similar.

Because of the second constraint \eqn{r constraint 2}, 
the Cardy states actually depends only on the sum of labels
\begin{eqnarray}
M \equiv \sum_i \mu_i M_i + 2 \sum_j \nu_j K_j R_j~ \in \bz_{2N}~,
\label{def M}
\end{eqnarray}
and thus we shall write them as $\ket{B;\{L_i\}, M ; \pm}$,
$\ket{B;\{L_i\}, M}^{(\sigma)}$ from here on.


It is now possible to count the numbers of compact D-branes 
and compare them with those of massless states
in the models with $N_M=1$, $1\leq N_L \leq 3$. 
See table 1. We note that some of the D-branes 
(cycles) do not have corresponding massless states and thus the number 
of D-brane exceed those of massless moduli.
This is because the ``would be" moduli 
which are non-normalizable in the non-compact geometry 
do not appear as massless states in the closed string spectrum.

\vskip1cm 

{\small
\hspace{-2cm}
\begin{tabular}{|c|c|c|c|}
 \hline
Models  & Geometric Identification &  No. of  massless  & 
No. of  basic \\ 
& & states &  vanishing cycles \\
\hline
 $M_{n-2} \otimes L_{n,1}$ & $ALE$ ($A_{n-1}$)   & $n-1$   & $n$   
   \\ \hline
 $M_{n-2}\otimes L'_{2n,n+2}$ & $CY_3$ ($A_{n-1}$)  & 
$\left\lb \frac{n-2}{2}\right\rb +1$
    & $n$   \\ \hline
 $M_{n-2}\otimes L_{n,n+1}$ & $CY_4$ ($A_{n-1}$)  & 0   &  $n$   
  \\ \hline
 $M_{n-2}\otimes L_{2n,1} \otimes L_{2n,1}$ &
$ALE$ ($A_{n-1}$) fibration over $\bc P^1$  &   $n-1$   & $2n$    \\ \hline
 $M_{n-2} \otimes L_{n\mu,K_1} \otimes L_{n\mu,K_2}$, 
& $ALE$ ($A_{n-1}$) fibration  &
    &    \\
$K_1+K_2=\mu$, 
$\mbox{G.C.D}\{K_i\}=1$
  & over $W \bc P^1 \lb K_1,K_2 \rb$  &  $n-1$  &  $ \mu n$  \\
 \hline
 $M_{n-2} \otimes L'_{4n,n+2} \otimes L'_{4n,n+2}$  &
$CY_3$ ($A_{n-1}$) fibration over $\bc P^1$  
&  $\left\lb \frac{n-2}{2}\right\rb +1 $  & $2n$    \\ \hline
 $M_{n-2} \otimes L'_{2n\mu,K_1(n+2)} \otimes L'_{2n\mu,K_2(n+2)}$, 
& $CY_3$ ($A_{n-1}$) fibration  &   &     \\
$K_1+K_2=\mu$, 
$\mbox{G.C.D}\{K_i\}=1$
  & over $W \bc P^1 \lb K_1,K_2 \rb$  
&  $\left\lb \frac{n-2}{2}\right\rb +1 $
  & $ \mu n$ \\
 \hline
 $M_{n-2}\otimes L_{3n,1} \otimes L_{3n,1} \otimes L_{3n,1} $ &
$ALE$ ($A_{n-1}$) fibration over $\bc P^2$  &   $n-1$   & $3n$    \\ 
\hline
 $M_{n-2} \otimes L_{n\mu,K_1} \otimes L_{n\mu,K_2}\otimes L_{n\mu,K_3}  $, 
& $ALE$ ($A_{n-1}$) fibration    &
   &     \\
$K_1+K_2+K_3=\mu$, 
$\mbox{G.C.D}\{K_i\}=1$
  & over $W \bc P^2 \lb K_1,K_2,K_3 \rb$  & $n-1$   & $ \mu n$  \\
 \hline

\end{tabular}
}

\begin{center}
Table 1
\end{center}

(``No. of  basic vanishing cycles'' means the number of 
compact BPS branes $\ket{B; L, M}$ with $L=0$. 
They are not necessarily homologically independent.
Generic cycles with $L \neq 0$ are expressed as 
superpositions of the basic ones.)

~


For the sake of later use let us 
derive the `charge integrality' condition for the R-sector.  
This is obtained from (\ref{U(1) integrality}) by a 1/2-spectral flow: 
first, $U(1)$-charges of $M_{K_{i}}$
and $L_{N_j,K_j}$-sectors are shifted under the flow as
\begin{eqnarray}
&&{m_i\over k_i+2}\, \Longrightarrow \,
 {1\over 2}+{m_i'\over k_i+2}~, \hskip1cm (m_i'\equiv m_i-1)\\
&&{K_jn_j\over N_j}\, \Longrightarrow \,
{1\over 2}+{K_jn_j'\over N_j}~, \hskip1.5cm (n_j'\equiv n_j+1)~.
\label{U(1)-charge R sector}
\end{eqnarray}
At the same time total $U(1)$-charge gets shifted by $\hat{c}/2$. 
Thus the condition for charge integrality in R-sector becomes
\begin{equation}
{m_i'\over k_i+2}+{K_jn_j'\over N_j}\in \bz+{\gamma\over 2}
 \label{recall}\end{equation} 
where $\gamma$ is defined as
\begin{equation}
\gamma=\hat{c}-(N_M+N_L)~.
\label{gamma}
\end{equation}
As it turns out, our models have different characteristics 
depending on the even-oddness of the parameter $\gamma$.

~

\noindent
{\bf A comment on the $\hat{c}=2$ case : }

Since we showed that the $(c,c)$ and $(a,a)$-type chiral primaries 
exist in the $\hat{c}=2$ case, one may suppose that 
the A-type compact branes also exist in $\hat{c}=2$ theory.  
However, this is not the case. In fact, the $(a,a)$-type 
chiral primary states
\eqn{anti-chiral state} contains in each $M_{k_i}$-sector  a state of the type 
\begin{eqnarray}
(\ell_i,-\ell_i)_L \otimes (k_i-\ell_i,\ell_i-k_i)_R.
\end{eqnarray}
Thus we cannot define the A-type Ishibashi
state except for the special case $\ell_i=k_i/2$.
Therefore, generic $\hat{c}=2$ theories
do not have sufficient number of
A-type Ishibashi states for constructing compact A-branes.

Strictly speaking, there is an exception; 
$L_{2,1}$ (with no $M_k$ factor), which 
corresponds to the Eguchi-Hanson space topologically equivalent with 
$T^*S^2$. In this case we can construct the A-type boundary state
$\ket{B;O}_A$ for compact brane as well as the B-type $\ket{B;O}_B$,
  both of which 
are associated to the $\cN=4$ massless character of $\ell=0$ \cite{ET}. 
This fact seems to contradict with the geometrical interpretation, 
since only one cycle $\cong S^2$ exists in 
the Eguchi-Hanson space. However, this apparent puzzle 
is resolved by analyzing cylinder amplitudes. 
The B-B and A-A overlaps become (in the NS sector)
\begin{eqnarray}
&&  \hskip-2cm \braB{B;O} e^{-\pi T H^{(c)}} \ketB{B;O} =
\braA{B;O} e^{-\pi T H^{(c)}} \ketA{B;O} =
\ch{\cN=4}{0}(\ell=0;it,0)~, ~(T\equiv 1/t)
\end{eqnarray}
as expected. 
Here $\ch{\cN=4}{0}(\ell=0;\tau,z)$ is the $\cN=4$ massless character
of $\ell=0$ \cite{ET}.  
On the other hand, the A-B overlap is evaluated as follows;
\begin{eqnarray}
\braA{B;O} e^{-\pi T H^{(c)}} \ketB{B;O} = 
\chi_{(-,+)}(p=i/2;it) - \int_0^{\infty} dp\, \frac{2}{\cosh \pi p}\,
 \chi_{(-,+)}(p; it)~, ~ (t \equiv 1/T)
\label{compact A-B}
\end{eqnarray}
where $\chi_{(-,+)}(p;\tau)$ is the twisted $\cN=2$ character
defined in \eqn{twisted massive}. Twisted character appears 
due to the difference in the boundary conditions.
We have used the fact that the absolute value squared of 
boundary wave function becomes  
\begin{eqnarray}
2 \sinh \pi p' \tanh \pi p' = 2\left(\cosh \pi p' - 
\frac{1}{\cosh \pi p'}\right)~,
\end{eqnarray}
and also used a contour deformation technique, which yields the first
term in \eqn{compact A-B}. 
Note that the second term in \eqn{compact A-B} 
appears with a negative sign, which means that the open channel 
amplitude includes negative norm states. This implies 
that compact A and B-branes are mutually incompatible,
and we discard the A-brane as in the other cases.
In conclusion our brane spectrum matches with
the geometrical expectation also in this case.

~

\subsection{Cylinder Amplitudes}

~

Now, let us analyze cylinder amplitudes ending on the compact BPS branes
in various models. 
We assume the parameter $\gamma$ \eqn{gamma}
to be an even integer $\gamma \in 2\bz$ for the time being. 
Calculation 
of various amplitudes becomes simplified under this assumption.

We start our analysis by working on the $\NS$-sector amplitudes;
\begin{eqnarray}
&& \hspace{-1cm} Z^{(\sNS)} (\{L_i\}, M | \{L_i'\}, M') (it) 
\equiv {}^{(\sNS)}\bra{B;\{L_i\},M} e^{-\pi T H^{(c)}} 
\ket{B;\{L_i'\},M'}^{(\sNS)},~
(T\equiv 1/t).
\label{NS overlap} 
\end{eqnarray}
The calculation is quite similar to the cylinder amplitudes for the 
B-branes in the compact Gepner models (see {\em e.g.} \cite{RS,BDLR}). 
The non-trivial point is the treatment of the projection operator
$P_{\msc{closed}}$. 
The following formulas are useful in  imposing 
the second constraint \eqn{r constraint 2};
\begin{eqnarray}
&& \sum_{\stackrel{a\in \bsz_{2(k+2)}}{L+a \in 2\bsz}}\,
e^{-2\pi i \frac{am}{2(k+2)} }\, \ch{(\sNS)}{L,a}
 \left(-\frac{1}{\tau}, \frac{z}{\tau}\right) \nn
&& \hspace{1cm}
= e^{i\pi \frac{k}{k+2}\frac{z^2}{\tau}}\sum_{\ell=0}^{k} 
\sin\left(\frac{\pi(L+1)(\ell+1)}{k+2}\right) \,
\left\lb \ch{(\sNS)}{\ell,m}(\tau,z) 
+ (-1)^L \ch{(\sNS)}{\ell,m+k+2}(\tau,z)\right\rb~.
\label{identity M k ch}
\end{eqnarray}
\begin{eqnarray}
&& \sum_{R\in \bsz_N}\, e^{2\pi i \frac{R K n}{N}} \,
\chi_0^{(\sNS)}\left(2KR;-\frac{1}{\tau}, \frac{z}{\tau}\right) \nn
&& \hspace{1cm}
= e^{i\pi \hat{c}_L \frac{z^2}{\tau}} \sqrt{\frac{N}{2K}} 
\sum_{w\in \bsz_{2K}}\,
\left\lb 
\int_0^{\infty}dp \, \frac{\sinh\left(\pi\sqrt{\frac{2K}{N}}p\right)
\sinh\left(\pi \sqrt\frac{2N}{K} p\right)}
{\left|\cosh \pi \left(\sqrt{\frac{N}{2K}}p
+i \frac{Kn+Nw}{2K} \right)\right|^2}
\chic^{(\sNS)}(p,Kn+Nw;\tau,z)
\right. \nn
&& \hspace{1cm}
\left. + \sum_{s=K+1}^{N+K-1}\, 2 \sin \left(\frac{\pi(s-K)}{N}\right)\,
\chid^{(\sNS)}(s,Kn+Nw;\tau,z)\right\rb~, ~(\hat{c}_L= 1+\frac{2K}{N})~.
\label{identity L ch}
\end{eqnarray}
We obtain (up to an overall normalization)
\begin{eqnarray}
&&Z^{(\sNS)}(\{L_i\}, M | \{L_i'\}, M') (it)
\propto  \sum_{\ell_i} \, 
\sum_{\stackrel{a_i \in \bsz_{2(k_i+2)}}{a_i\equiv \ell_i \, 
(\msc{mod}\, 2)}}\,
\sum_{a_j'\in \bsz_{N_j}} \, \sum_{a \in \bsz_N}\,
\sum_{r\in \bsz_{2N}}\,   \frac{1}{N} \cdot
\frac{1}{2N}   \nn
&& \hspace{2cm}\times \exp \left\lb
2\pi i \frac{r}{2N} \left\{
M'-M + \sum_i \mu_i a_i +2\sum_j \nu_j K_j a_j'
\right\}
\right\rb \nn
&& \hspace{2cm} \times \prod_i \prod_j \, N_{L_i,L_i'}^{\ell_i}\,
 \ch{(\sNS)}{\ell_i,a_i+2a}(it,0)\,
\chi_0^{(\sNS)}(2K_j(a_j'-a);it,0) \nn
&& =   \sum_{\ell_i} \, 
\sum_{\stackrel{a_i \in \bsz_{2(k_i+2)}}{a_i\equiv \ell_i \, 
(\msc{mod}\, 2)}}\,
\sum_{a_j'\in \bsz_{N_j}} \, \sum_{a \in \bsz_N}\,  \frac{1}{N} \,
 \delta^{(2N)}\left(M'-M + \sum_i \mu_i a_i +2\sum_j \nu_j K_j a_j'
\right)\nn
&& \hspace{2cm} \times  \prod_i \prod_j \,N_{L_i,L_i'}^{\ell_i}\, 
\ch{(\sNS)}{\ell_i,a_i+2a}(it,0)\,
\chi_0^{(\sNS)}(2K_j(a_j'-a);it,0)~.
\label{cylinder amplitude 1}
\end{eqnarray}
Here $N_{L_i,L_i'}^{\ell_i}$ are  the fusion coefficients of 
$SU(2)_{k_i}$;
\begin{eqnarray}
N_{L_i,L_i'}^{\ell_i}=
\left\{
\begin{array}{ll}
 1& ~~ |L_i-L_i'|\leq \ell_i \leq \min\lb L_i+L_i',\, 
2k_i-L_i-L_i'\rb~ \mbox{and }
       \ell_i \equiv |L_i-L_i'|~ (\mod\, 2)\\
 0& ~~ \mbox{otherwise}
\end{array}
\right. ~~.
\end{eqnarray}
The  integrality of total $U(1)$-charge is ensured by the $a$-summation, 
while the $a_i$ and $a_j'$ summations impose the constraint 
\eqn{r constraint 2} via the relations \eqn{identity M k ch}, 
\eqn{identity L ch}. Thanks to our assumption
\begin{eqnarray}
 2\sum_i \mu_i - 2\sum_j \nu_jK_j = -N\gamma \in 2N \bz~,
\end{eqnarray}
we may make the shifts $a_i\,\rightarrow\, a_i+2a$,
$a_j'\,\rightarrow\,a_j'-a$ in the factor $\delta^{(2N)}(\cdots)$. 
The $a$-summation is then decoupled and we obtain a simpler form
of amplitude
\begin{eqnarray}
&& Z^{(\sNS)}(\{L_i\}, M | \{L_i'\}, M') (it)
\propto   \sum_{\ell_i} \, 
\sum_{\stackrel{a_i \in \bsz_{2(k_i+2)}}{a_i\equiv \ell_i \, 
(\msc{mod}\, 2)}}\,
\sum_{a_j'\in \bsz_{N_j}} \, 
\delta^{(2N)}\left(M'-M + \sum_i \mu_i a_i +2\sum_j \nu_j K_j a_j'
\right) \nn
&& \hspace{1cm}
\times \, \prod_i \prod_j \,N_{L_i,L_i'}^{\ell_i}\,
 \ch{(\sNS)}{\ell_i,a_i}(it,0)\,
\chi_0^{(\sNS)}(2K_ja_j';it,0)~.
\label{cylinder amplitude 2}
\end{eqnarray}
In the cases of $\gamma \in 2\bz+1$ the $a$-summation is in general not 
decoupled and the calculation becomes  more complex. 


We next consider the open string Witten indices defined by 
\begin{eqnarray}
&& I(\{L_i\},M|\{L_i'\},M') = {}^{(\sR)}\bra{B;\{L_i\},M} 
e^{-\pi T H^{(c)}} e^{-i\pi J_0} 
\ket{B;\{L_i'\},M'}^{(\sR)} ~.
\label{osWI}
\end{eqnarray}
Under the assumption $\gamma \in 2\bz$, 
the $U(1)$-charge condition for R-sector
has the same form as \eqn{U(1) integrality}. 
In order to evaluate the amplitudes
we only have to replace the NS-characters 
$\ch{(\sNS)}{*,*}(it,0)$, $\chi_0^{(\sNS)}(*;it,0)$
by the $\tR$-characters, that is, 
the Witten indices (up to signs)
\begin{eqnarray}
&& \ch{(\stR)}{\ell_i,m_i}(it;0) = \delta^{(2(k_i+2))}(m_i-(\ell_i+1))
- \delta^{(2(k_i+2))}(m_i+(\ell_i+1))~,~\nn
&& 
\chi_0^{(\stR)}(2K_jr_j;it,0) = \delta^{(N_j)}\left(r_j-\frac{1}{2}\right)
- \delta^{(N_j)}\left(r_j+\frac{1}{2}\right)~,
\end{eqnarray}
and also replace the summation $\dsp \sum_{a\in \bsz_N}*$ with 
$\dsp \sum_{a \in \frac{1}{2}+\bsz_N}*$ because of 
the insertion of $e^{-i\pi J_0}$. 
We finally obtain 
\begin{eqnarray}
&& \hspace{-5mm}
I(\{L_i\}, M | \{L_i'\}, M') \propto
\sum_{\ell_i}\, \sum_{\al_i=\pm 1} \,
\sum_{\beta_j=\pm 1} \, \delta^{(2N)}\left(
M'-M + \sum_i \mu_i \al_i(\ell_i+1) + \sum_j \nu_j K_j \beta_j 
+ \frac{N}{2}\gamma \right) \nn
&& \hspace{5cm} \times \prod_i \prod_j \, N_{L_i,L_i'}^{\ell_i}\,
\sgn(\al_i) \sgn(\beta_j)~.
\label{osWI2} 
\end{eqnarray}  
The sector in which all of $L_i$ and $L_i'$ equal 0 play the basic role. 
Following  \cite{BDLR}, 
$\lb I_0\rb_{M,M'} \equiv I(\{L_i=0\},M| \{L_i'=0\},M')$ can be 
concisely expressed by introducing 
a cyclic operator $g$ defined by the action
\begin{eqnarray}
g~:~M ~\longmapsto~ M+2~,
\label{operator g} 
\end{eqnarray}
which satisfies $g^N=1$. 
Then we obtain from \eqn{osWI2}
\begin{eqnarray}
I_0 &=& \prod_i \prod_j\, (g^{\frac{\mu_i}{2}}-g^{-\frac{\mu_i}{2}} )\,
(g^{\frac{\nu_j K_j}{2}} - g^{-\frac{\nu_j K_j}{2}}) \, g^{\frac{N}{4}\gamma}
\nn
 &\propto & \prod_i \prod_j\, (1-g^{-\mu_i})\, (1-g^{\nu_j K_j})~,~~~
(\mbox{up to overall sign})~.
\label{osWI3} 
\end{eqnarray}
In this way we can derive 
the formula conjectured in \cite{Lerche}, which generalizes 
the one for the B-branes in the compact Gepner models \cite{RS,BDLR}.
We here emphasize that the formula \eqn{osWI3} is correct only in the case
with even $\gamma$. It is easy to check that, under the assumption 
$\gamma \in 2\bz$, the formula \eqn{osWI3} 
has the correct symmetry, namely\footnote
   {One way to confirm the result \eqn{symmetry I 0} 
    is to take the T-dual so that $L_{N_j,K_j}$-sectors 
     are realized as $\cN=2$ Liouville theories. 
     There, the compact branes are A-branes corresponding to 
     middle dimensional cycles.},
\begin{eqnarray}
&& I_0^t = I_0~, ~~~\mbox{for}~ \hat{c}=2,4~, \nn
&& I_0^t = -I_0 ~, ~~~\mbox{for}~ \hat{c}=3~.
\label{symmetry I 0}
\end{eqnarray} 

Let us next discuss the odd $\gamma$ cases that  
are somewhat more complicated.
We here present two examples.

~

\noindent
{\bf 1. $CY_3$ ($A_{n-1}$) : }

In this case we have $\gamma=3-2=1$. 
As was already discussed, 
we adopt half-integral values of the winding the number 
\begin{eqnarray}
w_1 \in \frac{1}{2}\bz_{4(n+2)}~, ~~~ n_1+2w_1 \in 2\bz~.
\label{w half-integer}
\end{eqnarray}
It is convenient to parameterize 
\begin{eqnarray}
&& k_1+2=N_1=N=n~, ~ K_1=\frac{n+2}{2}~, ~~~ (\mbox{for even $n$})~, \nn
&& k_1+2=n~,~ N_1=N=2n~,~ K_1=n+2~, ~~~ (\mbox{for odd $n$})~,
\end{eqnarray}
so that $K_1$ and $N_1/2$ (not $N_1$) are  relatively prime.

We first consider the even $n$ case. 
An important difference from the previous analysis is in  
the $U(1)$-charge condition for R-sector;
\begin{eqnarray}
\frac{m}{n}+ \frac{(n+2)n_1+2nw_1}{2n} \in \bz + \frac{\gamma}{2} = 
\bz+ \frac{1}{2}~, 
\end{eqnarray}
which leads to an extra insertion of $e^{i\pi a}$ in the amplitude.
Secondly, the ``$\dsp w\in \frac{1}{2}\bz$-rule'' \eqn{w half-integer}
gives rise to a replacement $a_1'\,\rightarrow\,2a_1'$.
We so obtain 
\begin{eqnarray}
&& \hspace{-1cm} I (L, M | L', M') 
\propto  \sum_{\ell} \, 
\sum_{\stackrel{a_1 \in \bsz_{2n}}{a_1\equiv \ell \, 
(\msc{mod}\, 2)}}\,
\sum_{a_1'\in \bsz_{n}} \, \sum_{a \in \bsz_n}\,
\sum_{r\in \bsz_{2n}}\,   \frac{1}{n} \cdot
\frac{1}{2n}   
\, \exp \left\lb
2\pi i \frac{r}{2n} \left\{
M'-M +  a_1 +(n+2) 2a_1'
\right\} \right\rb \nn
&& \hspace{3cm} \times  e^{i\pi a}\,  N_{L,L'}^{\ell}\,
 \ch{(\stR)}{\ell,a_1+2a}(it,0)\,
\chi_0^{(\stR)}((n+2)(2a_1'-a);it,0) ~.
\label{osWI CY3 1}
\end{eqnarray}
Since only the terms with $\dsp 
2a_1'-a =  \pm \frac{1}{2}$ contribute, 
we may replace the factor $e^{i\pi a}$ with
$e^{-i\frac{\pi}{2} \beta}$  ($\dsp \frac{\beta}{2}\equiv 2a_1'-a$, 
$\beta=\pm 1$), 
which cancels out the factor $\sgn(\beta)$ in \eqn{osWI2}.
The summation over $a$ is again decoupled.
We finally obtain the formula
\begin{eqnarray}
&& I(L, M | L', M') \propto
\sum_{\ell}\, \sum_{\al=\pm 1} \,
\sum_{\beta=\pm 1} \, \delta^{(2n)}\left(
M'-M + \al(\ell+1) + \beta \right) \, N_{L,L'}^{\ell}\,
\sgn(\al) ~.
\label{osWI CY3 2} 
\end{eqnarray}
In the $L=0$ sector, 
we have
\begin{eqnarray}
I_0\propto (1-g^{-1})(1+g) = g-g^{-1}~,
\label{osWI CY3 3}
\end{eqnarray}
that is anti-symmetric as is expected.

In the odd $n$ case the story becomes more complicated. 
Since the label of the Cardy states $R_1$ in $L'_{2n,n+2}$-sector  
runs over the range $R_1 \in \bz_{2n}$, it appears that
there exist twice as many  BPS branes. 
However, the proper Cardy states in $L'_{2n,n+2}$
have to be compatible with \eqn{w half-integer} and are given as  
$
\ket{R_1}^{(\sR)} + \ket{R_1+n}^{(\sR)}.
$
We thus still have the same number of compact branes, 
and one may restrict the label $M$ to the range $M\in \bz_{4n} \cap 2\bz$. 
Rewriting $M/2$ as $M$, we can obtain the same formula of Witten indices
\eqn{osWI CY3 2} (and \eqn{osWI CY3 3}).

~

\noindent
{\bf 2. $\hat{c}=4$, $M_{n-2}\otimes L'_{4n,n+2} \otimes L'_{4n,n+2}$ : }

This model has been identified as the $CY_3(A_{n-1})$-fibration over
$\bc P^1$ and we have $\gamma = 4-3=1$. 
We should again apply the 
``$\dsp w\in \frac{1}{4}\bz$ rule'' \eqn{m CY3 fiber} 
to the $L'_{4n,n+2}$-sectors, and make the replacements 
$a_i'\,\rightarrow\, 4a_i'$. 
A similar calculation leads to 
\begin{eqnarray}
&& I(L, M | L', M') \propto
\sum_{\ell}\, \sum_{\al=\pm 1} \,
\sum_{\beta_1=\pm 1}\sum_{\beta_2=\pm 1}
\nn
&& \hspace{2cm} \times  
\delta^{(4n)}\left(
M'-M + 2\al(\ell+1) + \beta_1 + \beta_2\right) \, N_{L,L'}^{\ell}\,
\sgn(\al) \sgn(\beta_1)~, 
\label{osWI CY4 1}
\end{eqnarray} 
and also in the $L=0$ sector,
\begin{eqnarray}
I_0\propto (1-g^{-2})(1-g)(1+g) = 2-g^2-g^{-2}~.
\label{osWI CY4 2}
\end{eqnarray}
Note that the factor $\sgn(\beta_2)$ was canceled 
by $e^{i\pi a}$ in the same way as the first example,
yielding a contribution $1+g$ in place of $1-g$ 
in \eqn{osWI CY4 2}.
This fact makes \eqn{osWI CY4 2} symmetric as should be.

~

\subsection{Comments on Non-compact BPS Branes}

~

We can similarly investigate aspects of non-compact BPS branes. 
The analysis is straightforward but more cumbersome technically.
We thus restrict ourselves to making some comments about non-trivial points.
We shall concentrate on the non-compact branes associated to 
the massive representations in each of $L_{N_j,K_j}$-sectors 
(`class 2' in the classification in \cite{ES-L}), 
and focus on the NS-sector.
 
The first fact which is in contrast to the case of compact branes is
that {\em both} the A and B-type boundary conditions are possible. 
This is because the $U(1)$-charges of massive representations 
are uncorrelated with their conformal weights. 
The desired Cardy states describing non-compact branes 
are now constructed in a similar manner 
to \eqn{Cardy ncg}: all we have to do is to take
the products of Cardy states in each sector and to let the 
projection $P_{\msc{closed}}$ act on them.

The Cardy states in each of the $M_{k_i}$-sectors 
are given as in \eqn{minimal Cardy states} 
(here we specify the A and B-type
boundary conditions by subscripts);
\begin{eqnarray}
&&\ket{L_i,M_i}_A = \sum_{\ell_i} \sum_{m_i} \, e^{i\pi \frac{\ell_i}{2}}
C_{L_i,M_i}(\ell_i,m_i)
\, \dket{\ell_i,m_i}_A ~, \label{minimal Cardy A} \\
&&\ket{L_i,M_i}_B = \sum_{\ell_i} \sum_{m_i} \, 
C_{L_i,M_i}(\ell_i,m_i)
\, \dket{\ell_i,m_i}_B ~. \label{minimal Cardy B}
\end{eqnarray}
Note that the A-B overlap of Ishibashi states yields 
the twisted minimal characters (see Appendix D), and 
the extra phase factor $e^{i\pi \frac{\ell_i}{2}}$
in \eqn{minimal Cardy A} is necessary for the consistency  with 
the modular bootstrap. (See the modular transformation formulas 
\eqn{modular twisted minimal}, and 
recall the fact that the identity brane $\ket{B;O}$ is a B-brane.)


The Cardy states in the $L_{N_j,K_j}$-sectors are more non-trivial. 
Since the branes are non-compact, we should allow the continuous 
spectrum of $U(1)$-charge in the open channel, while that in the closed 
channel should be still discrete. We thus have to 
introduce a one-parameter deformation of the extended massive 
characters \eqn{extended massive w}
\begin{eqnarray}
&& \hspace{-2cm}
\chic^{(\sNS)}(p,m;\tau,z,w) = q^{\frac{p^2}{2}} \Th{m}{NK}
\left(\tau,\frac{2z}{N}- \frac{w}{NK}\right) \,
\frac{\th_{3}(\tau,z)}{\eta(\tau)^3}~, ~~~ (w \in \br~, ~~0\leq w < 2NK)~,
\end{eqnarray}
and the associated Ishibashi states $\dket{p,m,\al}_A$,
$\dket{p,m,\al}_B$ ($\al \in \br$, $0\leq \al < 2NK$)
by the relations;
\begin{eqnarray}
\hspace{-1cm}
{}_A\dbra{p,m,\al} e^{-\pi T H^{(c)}} e^{2 \pi i z J_0 } \dket{p',m',\al'}_A
&=& {}_B\dbra{p,m,\al} e^{-\pi T H^{(c)}} e^{2 \pi i z J_0 } 
\dket{p',m',\al'}_B
\nn
&=& \delta(p-p') \delta_{m,m'}^{(2NK)} \chic^{(\sNS)}(p,m;\tau,z,\al'-\al)
~, ~~~(p,p'>0)
\nn
\hspace{-1cm}
{}_A\dbra{p,m,\al} e^{-\pi T H^{(c)}} e^{2\pi i z J_0 } \dket{p',m',\al'}_B
&=& {}_B\dbra{p,m,\al} e^{-\pi T H^{(c)}} e^{2 \pi i z J_0 } 
\dket{p',m',\al'}_A
\nn 
&=& \delta(p-p') \delta_{m,0}^{(2NK)} \delta_{m',0}^{(2NK)}\,
\chi_{(+,-)}(p;iT)~,
~~~(p,p'>0)
\end{eqnarray}
where $\chi_{(+,-)}(p;\tau)$ is the twisted massive character defined in
\eqn{twisted massive}.
By these definitions
the Ishibashi states $\dket{p,m,\al}_A$, $\dket{p,m,\al}_B$ are explicitly
constructed as the spectral flow sums of irreducible Ishibashi states, and 
the parameter $\al$ expresses the relative phase attached to each 
irreducible one. 

The desired pieces of A and B-type Cardy states in the $L_{N_j,K_j}$-sector 
are now given as follows \cite{RibS,ES-L,ASY,IPT2,FNP}:
they have different boundary wave functions; 
\begin{eqnarray}
&& \ket{P_j,\al_j}_B = \sqrt{\frac{2}{N_jK_j}} \int_0^{\infty}dp\, \sum_m\,
\cos(2\pi P_j p) f(p,m)\, \dket{p,m,\al_j}_B~, 
\label{Cardy massive B} \\
&& \ket{P_j,\al_j}_A = \frac{1}{\sqrt{2N_jK_j}} \int_0^{\infty}dp\, \sum_m\,
\left(e^{2\pi iP_jp}+ (-1)^m e^{-2\pi i P_j p}\right) 
f(p,m)\, \dket{p,m,\al_j}_A~, 
\label{Cardy massive A}
\end{eqnarray}
where we set 
\begin{eqnarray}
f(p,m) \equiv \frac{1}{\Psi_O^{(\sNS)}(-p,m)}
= \left(\frac{N^3}{2^3 K}\right)^{\frac{1}{4}}\, 
\frac
{\Gamma\left(-i\sqrt{\frac{2K}{N}}p\right) 
\Gamma\left(1-i \sqrt{\frac{2N}{K}}p\right)}
{\Gamma\left(\frac{1}{2}+\frac{m}{2K}-i \sqrt{\frac{N}{2K}}p\right)
\Gamma\left(\frac{1}{2}-\frac{m}{2K}-i \sqrt{\frac{N}{2K}}p\right)
}~,
\label{f p m}
\end{eqnarray}

~

We present a few comments:

~

\noindent
{\bf 1. }
The B-type Cardy state \eqn{Cardy massive B} is determined 
from the modular bootstrap, while the A-type \eqn{Cardy massive A} is not. 
This is because the identity brane is now a B-brane and hence 
the modular bootstrap is not powerful to determine the A-type Cardy 
state. Nevertheless, \eqn{Cardy massive A}, which is called `class $2'$'
in \cite{FNP}, has been constructed in
\cite{RibS,IPT2} as the `descent' of the $AdS_2$-brane \cite{BP} 
in the Euclidean $AdS_3$ \cite{PST}.\footnote
  {In the recent paper \cite{Hosomichi} the boundary wave function 
    of A-type \eqn{Cardy massive A} 
    has been also derived by the boundary bootstrap 
    approach based on the perturbative analysis in the dual 
     $\cN=2$ Liouville theory.} 
One can check that the difference of coefficients 
in \eqn{Cardy massive B} and \eqn{Cardy massive A} 
is consistent with reflection amplitudes of 
the $SL(2;\br)/U(1)$-coset model \cite{Teschner-reflection,GK}.

In the cigar models \eqn{Cardy massive B} would correspond to non-compact
D2-branes (partially wrapping on cigar) \cite{FNP}, 
while \eqn{Cardy massive A} do to non-compact D1-branes 
\cite{RibS,IPT2}. 
The continuous parameter $\al_j$ would express the angular positions 
and Wilson lines of D1, D2-branes respectively. 
However, one should keep it in mind that the backgrounds here have 
different geometries from the cigar due to the orbifolding procedure,
and thus the classical DBI analysis on the cigar as in \cite{RibS} 
cannot be simply applied to our case.


~

\noindent
{\bf 2. }
Computations of cylinder amplitudes are straightforward but
more complicated than the compact brane case. 
In the simplest example $M_{N-2}\otimes L_{N,1}$ 
some analysis has been already done in \cite{NST}\footnote
  {The analysis in \cite{NST} corresponds to the case with discrete $\al$.}.
It is important that all the open string amplitudes 
appearing in the A-A or B-B type overlaps are expanded 
by the products of minimal characters and extended massive 
characters with the {\em continuous\/} $U(1)$-charges
$\chic^{(*)}(p_j,\om_j;it,0)$, $\om_j \in \br$, $0\leq \om_j < 2N_jK_j$.
(See the modular transformation formula \eqn{S cont 2}.)

For the B-branes 
the projection $P_{\msc{closed}}$ acts in the same way as 
in the compact branes, namely imposes \eqn{U(1) integrality} and 
\eqn{r constraint 2}, and hence the B-branes 
can depend only on one parameter along the $U(1)$-direction.
As for the A-branes, on the other hand, $P_{\msc{closed}}$
only imposes the $U(1)$-charge integrality \eqn{U(1) integrality} 
and not the second constraint \eqn{r constraint 2}. 
If $K_j >1$, $P_{\msc{closed}}$ further gives rise to the summation 
over the shifts $\om_j\, \rightarrow\, \om_j +2N_j l_j$ 
($l_j \in \bz_{K_j}$) in the open channel character
$\chic^{(*)}(p_j,\om_j;it,0)$,  since the A-type Ishibashi states
$\dketc{p_j,m_j}_A$ exist only for $m_j= K_j n_j$ ($n_j \in
\bz_{2N_j}$). 

The A-B (or B-A)-type overlaps are expressed by the products of 
the twisted $\cN=2$ characters $\chi_{(-,+)}(p;it)$ \eqn{twisted
massive} and $\chi_{L\,(-,+)}(it)$ \eqn{twisted minimal}. 
We do not have spectral flow sums in the open channel in this case,
since only the $U(1)$-neutral states contribute to the twisted characters. 
The Cardy condition is expected to be satisfied with suitable 
spectral densities (after subtracting the IR divergences)
in all these cases.

~

\noindent
{\bf 3. }
We still have a possibility to construct other types of 
non-compact BPS branes based on the `class 3' boundary states
\cite{ES-L},  
which are associated with the massless matter representations 
in the $L_{N_j,K_j}$-sectors. 
In the cigar models 
it has been pointed out \cite{Eguchi-strings04,FNP} that 
they could describe the D2-branes found in \cite{RibS,IPT2} 
covering the whole cigar.  
However, as is suggested from a detailed analysis 
of cylinder amplitudes performed in \cite{FNP}, 
it seems difficult that these class 3 branes can satisfy the Cardy condition,
as far as we insist on {\em unitary representations} 
in the open string channel.
We leave this subtle problem to future works.

~


\subsection{Non-BPS Branes}

~

To close this section we discuss the Cardy states for non-BPS D-branes
that exhibit some interesting properties.  
To avoid unessential complexity we shall concentrate on
the simplest case of $N$ NS5 branes  (or $ALE(A_{N-1})$) described by
$M_{N-2} \otimes L_{N,1}$, 
and only consider the compact branes. Extensions to 
more general models and the cases of non-compact branes 
should be straightforward. 

The simplest non-BPS branes are of course obtained in the same way as  
in the flat backgrounds (see {\em e.g.} \cite{Sen-review}),
that is, by projecting out the RR-components of boundary states and 
by multiplying the remaining NSNS-components by $\sqrt{2}$.  
These branes are constructed using the descent
relation and the $\bz_2$-orbifolding 
acting by the space-time fermion number. 
Since we now possess the $\bz_N$-symmetries in the $U(1)$-charge 
sector, we may construct more non-trivial non-BPS branes 
 by using 
the $\bz_N$-orbifolding procedure. 
This type of non-BPS branes may be 
regarded as the natural extension of  the ``unstable B-branes'' in 
the $SU(2)$-WZW model presented in \cite{MMS}.

The basic prescription for their construction is summarized as follows
(we focus on the NSNS-sector for the time being);
\begin{enumerate}
 \item Start with the compact BPS brane \eqn{Cardy ncg}
\begin{eqnarray}
&& \ket{B;L,M}_B = \cN P_{\msc{closed}} \left\lb 
\ket{L,M}_{M_{N-2}} \otimes \ket{R=0}_{L_{N,1}}\right\rb~, ~~~\nn
&& \hspace{3cm}
(L+M\in 2\bz)~, 
\label{compact BPS NS5}
\end{eqnarray}
which includes the B-type Cardy states for both sectors.
Consider the ``wrong-dimensional'' BPS branes $\ket{B;L,M}'_{AB}$
which is defined by reversing formally the boundary condition 
in the $M_{N-2}$-sector of \eqn{compact BPS NS5}.
The subscript $AB$ indicates to take the A and B-type
boundary conditions for the $M_{N-2}$ 
and $L_{N,1}$-sectors, respectively. 
 \item Sum up $\ket{B;L,M}'_{AB}$ over the spectral flows in the 
open string channel, namely, we define
\begin{eqnarray}
&& \ket{B;L}_{AB} = \sum_{r\in \bsz_N}\, \ket{B;L,L+2r}'_{AB}~, 
\label{nbps compact} 
\end{eqnarray}
which should be the desired boundary states of non-BPS branes.  
\end{enumerate} 
It is important to note that, although the `wrong BPS brane' 
$\ket{B;L,M}'_{AB}$ is not compatible with the charge-integrality condition, 
\eqn{nbps compact}
is consistent because the sum over $r\in \bz_N$ projects out 
states with fractional $U(1)$-charges. They actually consist of 
Ishibashi states with integral $U(1)$-charges {\em separately\/}
in each sector, $M_{N-2}$ and $L_{N,1}$. 
They do not satisfy the $\cN=2$ boundary conditions \eqn{A-type}
or \eqn{B-type},
and at most preserve the $\cN=1$ superconformal symmetry.

The modular bootstrap relation characterizing 
this brane is written as 
\begin{eqnarray}
&& {}_B \bra{B;O} e^{-\pi T H^{(c)}} \ket{B;L}_{AB} 
= \chi_{L\,(-+)}(it) \, \widehat{\chig}(it,0)~,
\label{mb nbps 1} 
\end{eqnarray}
which fixes the overall normalizations in  \eqn{nbps compact}.
Here $\chi_{L\,(-+)}(\tau)$ is the twisted minimal character 
in $M_{N-2}$-sector defined in \eqn{twisted minimal}, and  
we have introduced a function 
\begin{eqnarray}
&& \hspace{-5mm}
\widehat{\chig}(\tau,z) \equiv \sum_{r\in \bsz_{N}}\, \chig(2r;\tau,z) 
= q^{-\frac{1}{4N}}\sum_{n\in \bsz}\,
\frac{(1-q)q^{\frac{n^2}{N}+n-\frac{1}{2}} 
e^{2\pi i z \left(\frac{2}{N}n+1\right)}}
{\left(1+e^{2\pi i z}q^{n+\frac{1}{2}}\right)
\left(1+e^{2\pi i z}q^{n-\frac{1}{2}}\right)}\,
\frac{\th_3(\tau,z)}{\eta(\tau)^3}~, 
\label{chig hat} 
\end{eqnarray}
which is  
the sum of irreducible graviton character 
over the whole spectral flows. 


Note that \eqn{nbps compact}
is naturally regarded as the $\bz_N$-extension of the non-BPS branes in 
the flat background \cite{Sen-review};
\begin{eqnarray}
\ket{Dp}_{\msc{non-BPS}} = \frac{1}{\sqrt{2}} \left(\ket{Dp}'+ 
\overline{\ket{Dp}}'\right)~,
\label{nbps brane flat}
\end{eqnarray}
where $\ket{Dp}'$ ($\overline{\ket{Dp}}'$) expresses the boundary state
for the `wrong-dimensional' BPS (anti-)$Dp$-brane.
Since each term of R.H.S is not compatible with the GSO condition,
the boundary state $\ket{Dp}_{\msc{non-BPS}}$ cannot be decomposed into
constituent branes. In the same sense our non-BPS brane 
\eqn{nbps compact}
is irreducible and not decomposable to constituent  
boundary states. 

It is straightforward to work out cylinder amplitudes 
and one can show that Cardy condition is always satisfied
for the non-BPS brane \eqn{nbps compact}. 
Namely, all the overlaps are interpreted as 
correct open string one-loop amplitudes. 
The overlaps with compact BPS branes are easy to evaluate 
in a similar manner as \eqn{mb nbps 1}, and we also find 
\begin{eqnarray}
{}_{AB}\bra{B;L_1} e^{-\pi T H^{(c)}} \ket{B;L_2}_{AB}
= \sum_L N^L_{L_1,L_2}\, \widehat{\ch{}{L}}(it,0) \, 
\widehat{\chig}(it,0)~,
\label{overlap nbps}
\end{eqnarray}
where $\widehat{\chig}(it,0)$ is defined in \eqn{chig hat},
and also we set
\begin{eqnarray}
\widehat{\ch{}{\ell}}(\tau,z) \equiv \sum_{r \in \bsz_{N}}\,
\ch{(\sNS)}{\ell,\ell+2r}(\tau,z) = 
\sum_{s\in \bsz_{N-2}}\, c^{(N-2)}_{\ell,\ell+2s}(\tau)\,
\Th{2\ell + 2N s}{N-2}\left(\frac{\tau}{2N}, \frac{z}{N}\right)~.
\label{ch hat}
\end{eqnarray}
Here $c^{(k)}_{\ell,m}(\tau)$ denotes the level $k$ string function of
$SU(2)$.
Note that the open string channel includes states with fractional 
$U(1)$-charges even in the self-overlap cases, 
suggesting that this boundary state really 
describes non-BPS D-branes.

To clarify the non-BPS nature it is important to examine the RR-sectors
of boundary states. 
It is easy to construct the RR-counterpart of \eqn{nbps compact}.
However, we have to take account of 
the compatibility with the charge integrality 
and GSO projection together with the Minkowski part $\br^{5,1}$.
Let us now recall the formula for  $U(1)$-charges in the R-sector 
(\ref{U(1)-charge R sector}) ,
\begin{eqnarray}
&&M_{N-2} \mbox{-sector}: ~ {1\over 2}+{m'\over N},\hskip1.5cm 
L_{N,1} \mbox{-sector}: ~ {1\over 2}+{n' \over N}~.
\end{eqnarray}
As we noted above, the sum over $r$ in (\ref{nbps compact}) forces 
the fractional parts of the $U(1)$-charges $m'/N,n'/N$ to vanish and hence
$U(1)$-charge becomes $1/2$ in each R-sector.

When the boundary condition is flipped from B to A-type 
in the $M_{N-2}$-sector, 
$U(1)$-charge of the right mover
changes from $1/2$ to $-1/2$ and there is a net change of $U(1)$-charge 
by $1$. Then a compensating change 
must happen in the flat sector along $\br^{5,1}$.
 One has to shift the dimension of the brane by one and obtains a brane with a wrong dimension
$|Dp'\rangle'$ ($p'$ is even(odd) for type IIA (IIB) string theory). 
Note that in the NS5-brane background a BPS brane has odd dimensions 
extended in the direction transverse to NS5-brane
and thus has odd (even) dimensions 
extended along the NS5-branes in type IIA (IIB) theory.
Non-BPS brane is then
given by
\begin{eqnarray}
&& \ket{Dp'}^{'(\sNS)} \otimes \ket{B;L}_{AB}^{(\sNS)}
\pm  \ket{Dp'}^{'(\sR)} \otimes \ket{B;L}_{AB}^{(\sR)}
~.
\label{nbps 2nd type}
\end{eqnarray}

In the present example it is also possible to construct a non-BPS brane 
of the type known in the flat space-time
\begin{eqnarray}
&& \sqrt{2} \ket{Dp}^{(\sNS)} \otimes \ket{B;L}_{AB}^{(\sNS)}~.
\label{nbps 1st type}
\end{eqnarray}
Here $p$ is odd (even) in the type IIA (IIB) theory. 
The second one \eqn{nbps 1st type} describes 
an overall wrong-dimensional brane without RR-component and 
the overall factor $\sqrt{2}$ is necessary by the same reason as 
in the flat case. 
The first one \eqn{nbps 2nd type} is more interesting and
characteristic for this conformal system. It is an overall {\em correct\/}
dimensional brane 
and has a non-vanishing RR-component. 
However, the boundary wave functions for the  RR-ground states 
always vanish, implying they have no RR-charges 
(vanishing periods in other words). 
In fact, as is obvious from the construction, they have the vanishing 
$\bz_N$-brane charge valued in the twisted K-group
(see {\em e.g.} \cite{twisted K}).
This type of branes also break the space-time SUSY completely, 
and one can show 
that their self-overlaps always include open string tachyons. 
In fact they always contain a contribution of $M_{N-2}$-sector
\begin{eqnarray}
\frac{1}{2} 
\left(\ch{(\sNS)}{0,2}(it,0)- \ch{(\stNS)}{0,2}(it,0)\right)
\equiv \frac{1}{2} 
\left(\ch{(\sNS)}{N-2,N-2}(it,0)+\ch{(\stNS)}{N-2,N-2}(it,0)\right)~,
\end{eqnarray}
in the open string amplitudes, where
the second term originates from the RR-boundary states. 
In the second equality we used the formula \eqn{field identification}.
It yields the leading IR behavior;~  $\sim e^{-2\pi t (h-1/2)}$
with 
\begin{eqnarray}
h= \frac{1}{2} - \frac{1}{N}~. 
\label{tachyon nbps compact}
\end{eqnarray}  
We have thus found open string tachyon modes
for any value $N\geq 2$. 
We also remark that the involution $L\,\rightarrow\, N-2-L$ 
flips the sign in front of the RR-component in \eqn{nbps 2nd type}
as in the BPS branes.
Hence it is actually enough to only consider the plus sign in 
\eqn{nbps 2nd type}.

~

We finally make a few comments:

~

\noindent
{\bf 1.}
In the recent paper \cite{Kutasov2} 
D-brane configurations in NS5-backgrounds breaking/not breaking
the space-time SUSY have been investigated in detail 
by means of the DBI action and an interesting geometrical interpretation of 
tachyon condensation in the non-BPS branes has been proposed 
in the Little String Theory (LST) \cite{GK}. 
Among other things, it is claimed that the BPS branes 
lying along the NS5 and breaking the space-time
SUSY completely, should be interpreted as the non-BPS branes in LST. 
It will be interesting to compare this type of branes with our boundary states 
for the compact non-BPS branes of the first type 
\eqn{nbps 2nd type}. It is shown in \cite{Kutasov2} that 
the open string modes describing the positions of these D-branes
transverse to NS5 become tachyonic and have the mass squared\footnote
   {Although the $S^1$-compactification is taken in \cite{Kutasov2},
    the tachyon mass given there
   does not depend on the compactification radius. }
\begin{eqnarray}
\al' M^2_T = -\frac{1}{N}~.
\end{eqnarray}
This coincides precisely with our calculation of tachyon mass
\eqn{tachyon nbps compact}.

~

\noindent
{\bf 2. }
We can also consider extensions 
to more general models and also the case of non-compact branes. 
The consistency with the GSO projection
in RR-sector is the only non-trivial point. For instance, let us consider 
the models with $N_M=1$, $N_L \geq 1$, which are identified with 
the ALE fibrations as we addressed before, and focus on the compact 
non-BPS branes. 
We then find 
\begin{itemize}
 \item If $\gamma (\equiv \hat{c}-N_M-N_L)$ is even, 
we have a similar situation as 
in the NS5 case above. 
  Namely, we have two types of non-BPS branes with/without
  the RR-components such as \eqn{nbps 2nd type}, 
  \eqn{nbps 1st type}.
 \item If $\gamma$ is odd, the boundary states of the type \eqn{nbps 2nd
       type} is not allowed since the charge integrality condition in the R-sector (\ref{recall})
       is not satisfied
       (recall that L.H.S of (\ref{recall}) vanishes due to $r$-summation) 
       and R-sector 
       does not appear in the boundary state.
       We have only the non-BPS branes without RR-components as 
       in the flat background. 
\end{itemize}

~



\section{Conclusions}

~

In this paper we have discussed various aspects of  
string theory compactification on non-compact Calabi-Yau manifolds
by making use of $SL(2;\br)/U(1)$ supercoset theories 
coupled to ${\cal N}=2$ minimal models.
We have used the extended characters of ${\cal N}=2$ SCA for 
the $SL(2;\br)/U(1)$ theory together with 
the irreducible characters for the minimal model
and determined the closed string massless spectrum, open string Witten index 
and (non-) BPS boundary states. 

Important aspect of the massless spectrum is the following: 
at our non-compact Gepner points
Calabi-Yau 3 and 4-folds possess only $(a,c)$ or $(c,a)$-type 
massless states,  
while $(c,c)$ or $(a,a)$-type states are absent in the spectrum.
Thus the theory possesses only K\"ahler structure deformations. 
In the T-dual ${\cal N}=2$ Liouville description 
the theory possesses only complex structure deformations 
corresponding to the special Lagrangian cycles. 

This is the characteristic feature 
of the space-time with a conifold singularity and thus our models 
describe generalized conifold singularities in CY 3 and 4-folds:
${\cal N}=2$ Liouville theories 
describe their deformations while $SL(2;\br)/U(1)$ theories
describe their resolutions.

On the other hand, in the case of K3 surface our models 
possesses equal numbers of $(a,c)$, $(c,a)$, $(a,a)$ and $(c,c)$ states, 
which corresponds to the characteristic feature 
of ADE type (hyperK\"{a}hler) singularities.

We have also studied the spectrum of D-branes in our non-compact models:
only the B-type branes are allowed as compact branes (or 
only the A-branes are possible in the $\cN=2$ Liouville theory)
and the open string Hilbert space describing compact branes
consists of extended graviton representations of $SL(2;\br)/U(1)$-sector and  
representations of the minimal sector. 
We have compared the spectra of compact branes  
with those of the massless states in the closed string sector.  
Some of the BPS branes (homology cycles) are 
not associated with massless states:
corresponding space-time fields are frozen 
due to non-normalizability of the wave function.
We have also seen that the 
cylinder amplitudes of
$\tR$-sector reproduce expected 
intersection numbers of vanishing cycles.

The Cardy states describing non-BPS D-branes are also discussed.
They are expressed by boundary states breaking the $\cN=2$ 
superconformal symmetry and identified as 
extensions of the ``unstable B-branes" in the $SU(2)$-WZW model.
They may possess (massive) RR-components contrary to non-BPS branes in the 
flat background. 

Geometry of the deformed side of the conifold is relatively easy to study 
due to the absence of quantum corrections and one can
study the Lagrangian cycles using Liouville theory.
On the other hand, possible quantum corrections make 
the geometry of resolved conifold difficult to understand. 
Even the dimensionality of the branes may not be well-defined.
It is interesting to see if the description 
by means of the $SL(2;\br)/U(1)$ theory will help
our understanding of the geometry of resolved conifold.


\section*{Acknowledgments}
\indent

Y. S. would like to thank N. Ishibashi, Y. Nakayama and Y. Satoh
for valuable comments and discussions.

The research of T. E. and Y. S. is partially  supported by 
Japanese Ministry of Education, 
Culture, Sports, Science and Technology.


\newpage
\appendix
\noindent
{\huge \bf  Appendix}

\section{Notations}


~

We here summarize our notations of theta functions.
We set  $q\equiv e^{2\pi i \tau}$ and  $y\equiv e^{2\pi i z}$, 
 \begin{eqnarray}
&& \theta_1(\tau,z) =i\sum_{n=-\infty}^{\infty}(-1)^n q^{(n-1/2)^2/2} y^{n-1/2}
  \equiv 2 \sin(\pi z)q^{1/8}\prod_{m=1}^{\infty}
    (1-q^m)(1-yq^m)(1-y^{-1}q^m), \nn
&&  \theta_2(\tau,z)=\sum_{n=-\infty}^{\infty} q^{(n-1/2)^2/2} y^{n-1/2}
  \equiv 2 \cos(\pi z)q^{1/8}\prod_{m=1}^{\infty}
    (1-q^m)(1+yq^m)(1+y^{-1}q^m)~, \nn
&& \theta_3(\tau,z)=\sum_{n=-\infty}^{\infty} q^{n^2/2} y^{n}
  \equiv \prod_{m=1}^{\infty}
    (1-q^m)(1+yq^{m-1/2})(1+y^{-1}q^{m-1/2})~, \nn
&& \theta_4(\tau,z)=\sum_{n=-\infty}^{\infty}(-1)^n q^{n^2/2} y^{n}
  \equiv \prod_{m=1}^{\infty}
    (1-q^m)(1-yq^{m-1/2})(1-y^{-1}q^{m-1/2}) ~,
\end{eqnarray}
 \begin{eqnarray}
 \Th{m}{k}(\tau,z)&=&\sum_{n=-\infty}^{\infty}
 q^{k(n+\frac{m}{2k})^2}y^{k(n+\frac{m}{2k})} ~.
\end{eqnarray}
 \begin{equation}
 \eta(\tau)=q^{1/24}\prod_{n=1}^{\infty}(1-q^n)~.
 \end{equation}

~

\section{Character Formulas of $\cN=2$ Minimal Model}


~

The character formulas of 
the level $k$ $\cN=2$ minimal model $(\hat{c}=k/(k+2))$ are described 
as the branching functions of 
the Kazama-Suzuki coset $\dsp \frac{SU(2)_k\times U(1)_2}{U(1)_{k+2}}$
defined by
\begin{eqnarray}
&& \chi_{\ell}^{(k)}(\tau,w)\Th{s}{2}(\tau,w-z)
=\sum_{\stackrel{m\in \bsz_{2(k+2)}}{\ell+m+s\in 2\bsz}} \chi_m^{\ell,s}
(\tau,z)\Th{m}{k+2}(\tau,w-2z/(k+2))~, \nn
&& \chi^{\ell,s}_m(\tau,z) \equiv  0~, ~~~ \mbox{for $\ell+m+s \in 2\bz+1$}~,
\label{branching minimal}
\end{eqnarray}
where $\chi_{\ell}^{(k)}(\tau,z)$ is the spin $\ell/2$ character of 
$SU(2)_k$;
\begin{eqnarray}
&&\chi^{(k)}_{\ell}(\tau, z) 
=\frac{\Th{\ell+1}{k+2}(\tau,z)-\Th{-\ell-1}{k+2}(\tau,z)}
                        {\Th{1}{2}(\tau,z)-\Th{-1}{2}(\tau,z)}
\equiv \sum_{m \in \bsz_{2k}}\, c^{(k)}_{\ell,m}(\tau)\Th{m}{k}(\tau,z)~.
\label{SU(2) character}
\end{eqnarray}
The branching function $\chi^{\ell,s}_m(\tau,z)$ 
is explicitly calculated as follows (see, {\em e.g.} \cite{KYY});
\begin{equation}
\chi_m^{\ell,s}(\tau,z)=\sum_{r\in \bsz_k}c^{(k)}_{\ell, m-s+4r}(\tau)
\Th{2m+(k+2)(-s+4r)}{2k(k+2)}(\tau,z/(k+2))~.
\end{equation}
Then, the character formulas of unitary representations 
are written as 
\begin{eqnarray}
&& \hspace{-1cm} \ch{(\sNS)}{\ell,m}(\tau,z) = \chi^{\ell,0}_m(\tau,z)
+\chi^{\ell,2}_m(\tau,z)~, \nn
&&  \hspace{-1cm} \ch{(\stNS)}{\ell,m}(\tau,z) = \chi^{\ell,0}_m(\tau,z)
-\chi^{\ell,2}_m(\tau,z)\equiv 
e^{-i\pi\frac{m}{k+2}}\ch{(\sNS)}{\ell,m}\left(\tau,z+\frac{1}{2}\right)~, \nn
&& \hspace{-1cm} \ch{(\sR)}{\ell,m}(\tau,z) = \chi^{\ell,1}_m(\tau,z)
+\chi^{\ell,3}_m(\tau,z) \equiv 
q^{\frac{k}{8(k+2)}}y^{\frac{k}{2(k+2)}}
\ch{(\sNS)}{\ell,m+1}\left(\tau,z+\frac{\tau}{2}\right)~, \nn
&&  \hspace{-1cm} \ch{(\stR)}{\ell,m}(\tau,z) = \chi^{\ell,1}_m(\tau,z)
-\chi^{\ell,3}_m(\tau,z) \equiv
- e^{-i\pi\frac{m+1}{k+2}}q^{\frac{k}{8(k+2)}}y^{\frac{k}{2(k+2)}}
\ch{(\sNS)}{\ell,m+1}\left(\tau,z+\frac{1}{2}+\frac{\tau}{2}\right)~. 
\label{minimal character}
\end{eqnarray}
By definition, we may restrict to $\ell+m \in 2\bz$ for the  $\NS$ and
$\tNS$ sectors, and to $\ell+m \in 2\bz+1$ for the $\R$ and $\tR$
sectors.  It is convenient to define 
$\ch{(\sigma)}{*}(\tau,z)\equiv 0$ unless these conditions for $\ell$,
$m$ are satisfied.
Note that the character identity (``field identification'') holds
\begin{eqnarray}
&& \chi^{k-\ell,s+2}_{m+k+2}(\tau,z) = \chi^{\ell,s}_m(\tau,z)~,
\end{eqnarray}
or equivalently, 
\begin{eqnarray}
&& \ch{(\sigma)}{k-\ell,m+k+2}(\tau,z)= \ch{(\sigma)}{\ell,m}(\tau,z)
~,~~(\sigma = \NS,\, \R)~,\nn
&& \ch{(\sigma)}{k-\ell,m+k+2}(\tau,z)= -\ch{(\sigma)}{\ell,m}(\tau,z)
~,~~(\sigma = \tNS,\, \tR)~. 
\label{field identification}
\end{eqnarray}

We next present the modular transformation formulas.
To this aim it is convenient to introduce the notations 
\begin{eqnarray}
&& T\cdot \NS = \tNS~,~~ T\cdot \tNS = \NS~, ~~ T\cdot \R = \R ~, ~~
 T\cdot \tR = \tR~, 
\label{T sigma} \\
&& S\cdot \NS = \NS~,~~ S\cdot \tNS = \R~, ~~ S\cdot \R = \tNS ~, ~~
 S\cdot \tR = \tR~.
\label{S sigma} 
\end{eqnarray}
We also define
\begin{eqnarray}
\kappa(\sigma)=
\left\{
\begin{array}{ll}
 1 & ~~ \sigma=\NS,\,\tNS,\, \R  \\
 -i & ~~ \sigma=\tR
\end{array}
\right. ~, ~~~~
\nu(\sigma)=
\left\{
\begin{array}{ll}
 0 & ~~ \sigma=\NS,\,\tNS\\
 1 & ~~ \sigma=\R,\, \tR
\end{array}
\right. ~~.
\label{kappa sigma}
\end{eqnarray}

Modular transformations are given by
\begin{eqnarray}
&& \chi^{\ell,s}_m(\tau+1,z) = e^{2\pi i \left(h^{(\sNS)}(\ell,m)+ 
\frac{s^2}{8}-\frac{k}{8(k+2)}\right)}\,  \chi^{\ell,s}_m(\tau,z)~,
\label{T chi l s m} \\
&& \chi^{\ell,s}_m\left(-\frac{1}{\tau},\frac{z}{\tau}\right)=
e^{i\pi \frac{k}{k+2}\frac{z^2}{\tau}}\,
\sum_{\ell'=0}^k \,\sum_{m\in \bsz_{2(k+2)}}\, \sum_{s\in \bsz_4}\,
S^{\ell',m'}_{\ell,m}\, \frac{1}{2} e^{-i \pi \frac{ss'}{2}}\,
\chi^{\ell',s'}_{m'}(\tau,z)~, 
\label{S chi l s m}
\end{eqnarray} 
where $\dsp h^{(\sNS)}(\ell,m) \equiv
\frac{\ell(\ell+2)-m^2}{4(k+2)}$ and 
$S^{\ell',m'}_{\ell,m}$ is the modular coefficients
\begin{eqnarray}
S^{\ell',m'}_{\ell,m} = 
\sqrt{\frac{2}{k+2}}\sin\left(\frac{\pi(\ell+1)(\ell'+1)}{k+2}\right)
\cdot \frac{1}{\sqrt{2(k+2)}}e^{2\pi i \frac{mm'}{2(k+2)}} ~.
\label{minimal S}
\end{eqnarray}
Equivalently, 
\begin{eqnarray}
&& \ch{(\sigma)}{\ell,m}(\tau+1,z) = 
e^{2\pi i \left(h^{(\sigma)}-\frac{k}{8(k+2)}\right)}\,
\ch{(T\cdot \sigma)}{\ell,m}(\tau,z)~,
\label{T ch} \\
&& \ch{(\sigma)}{\ell,m}\left(-\frac{1}{\tau}, \frac{z}{\tau}\right)
= \kappa(\sigma) e^{i\pi \frac{k}{k+2}\frac{z^2}{\tau}}\,
\sum_{\ell'=0}^k \,\sum_{m\in \bsz_{2(k+2)}}\, 
S^{\ell',m'}_{\ell,m}\, \ch{(S\cdot \sigma)}{\ell',m'}(\tau,z)~,
\label{S ch}
\end{eqnarray}
where $h^{(\stNS)}(\ell, m)= h^{(\sNS)}(\ell,m)$,
$h^{(\sR)}(\ell,m)= h^{(\stR)}(\ell,m) = h^{(\sNS)}(\ell,m) + 1/8$.
We note the spectral flow relations (${}^{\forall}a, {}^{\forall}b\in \bz$);
\begin{eqnarray}
q^{\frac{k}{2(k+2)}a^2} e^{2\pi i \frac{k}{k+2} a z}\,
\chi^{\ell,s}_m(\tau,z+a\tau+b) = 
e^{-i \pi s b} e^{2\pi i \frac{m}{k+2} b} \,
\chi^{\ell,s-2a}_{m-2a}(\tau,z)~.
\label{chi l s m spectral flow}
\end{eqnarray}
Especially, they have the (quasi-)periodicity of period $k+2$
with respect to $z$.   
We also note the formula of Witten index
\begin{eqnarray}
\lim_{z\,\rightarrow\,0}\,\ch{(\stR)}{\ell,m}(\tau,z) =
\delta^{(2(k+2))}_{m,\ell+1} - \delta^{(2(k+2))}_{m,-(\ell+1)}~.
\label{WI minimal}
\end{eqnarray}

~

\section{Extended Characters and Their Modular Properties}

~

In this Appendix we summarize useful properties of the extended
characters \eqn{chi c}, \eqn{chi d} and \eqn{chi 0}
in $L_{N,K}$-sector, {\em i.e.} the $SL(2;\br)/U(1)$
supercoset with $k=N/K$ ($N,K\in \bz_{>0}$) introduced in
\cite{ES-L,ES-BH}. 
They are explicitly written as 
\begin{eqnarray}
&& \hspace{-1cm} \chic^{(\sNS)}(p,m;\tau,z) 
= q^{\frac{p^2}{2}} \Th{m}{NK}\left(\tau,\frac{2z}{N}\right)\,
  \frac{\th_3(\tau,z)}{\eta(\tau)^3}~,
\label{chi c explicit} \\
&& \hspace{-1cm}
\chid^{(\sNS)}(s,s+2Kr;\tau,z) 
= \sum_{n\in \bsz}\, \frac{\left(yq^{N\left(n+\frac{2r+1}{2N}\right)}
\right)^{\frac{s-K}{N}}
y^{2K\left(n+\frac{2r+1}{2N}\right)} q^{NK\left(n+\frac{2r+1}{2N}\right)^2}
}{1+yq^{N\left(n+\frac{2r+1}{2N}\right)}}\,
\frac{\theta_3(\tau,z)}{\eta(\tau)^3}~, 
\label{chi d explicit} \\
&& \hspace{-1cm} 
\chid^{(\sNS)}(s,m;\tau,z) \equiv 0~, ~~~(m-s \not\in 2K\bz)~, \nn
&&
\hspace{-1cm}
\chi_0^{(\sNS)}(2Kr;\tau,z)= q^{-\frac{K}{4N}}\,\sum_{n\in\bsz}\, 
\frac{
(1-q)
q^{NK\left(n+\frac{r}{N}\right)^2+N\left(n+\frac{2r-1}{2N}\right)} 
y^{2K\left(n+\frac{r}{N}\right)+1}
}{\left(1+yq^{N\left(n+\frac{2r+1}{2N}\right)}\right)
\left(1+yq^{N\left(n+\frac{2r-1}{2N}\right)}\right)}
\, \frac{\theta_3(\tau,z)}{\eta(\tau)^3}~. 
\label{chi 0 explicit} \\
&& \hspace{-1cm} \chi_0^{(\sNS)}(m;\tau,z) \equiv 0~, ~~~ (m\not\in 2K\bz)~.
\nonumber
\end{eqnarray}
Expressions for other spin structures are readily derived 
from the definitions \eqn{extended os}.


The formulas of Witten indices are given as 
\begin{eqnarray}
\hspace{-5mm}
\chic^{(\stR)}(p,m;\tau,0)=0~, ~~~
\chid^{(\stR)}(s,m;\tau,0) = -\delta^{(2NK)}_{m,s-K}~,~~~
\chi_0^{(\stR)}(m;\tau,0) = \delta^{(2NK)}_{m,K}- \delta^{(2NK)}_{m,-K}~.
\label{Witten index}
\end{eqnarray}
The following relation of spectral flow is obvious from 
the definitions (${}^{\forall}a, b \in \bz$);
\begin{eqnarray}
&&q^{\frac{\hat{c}}{2}a^2}e^{2\pi i \hat{c} a z} \, 
\chi_*^{(\sigma)}(*,m;\tau,z+a\tau+b) 
=\epsilon_{\sigma}(a,b)e^{2\pi i \frac{m}{N} b} \chi_*^{(\sigma)}(*,m+2Ka;\tau,z)~,
\label{chi spectral flow relation}
\end{eqnarray}
where the sign factor $\epsilon_{\sigma}(a,b)$ 
is given by $\epsilon_{\sNS}(a,b)=1$,
$\epsilon_{\stNS}(a,b)=(-1)^a$, $\epsilon_{\sR}(a,b)=(-1)^{b}$, 
$\epsilon_{\stR}(a,b)=(-1)^{a+b}$,
respectively. 
Especially, the extended characters have the (quasi-)periodicity 
with period $N$ under the spectral flows. 
We also note the ``charge conjugation relations'' of extended characters;
\begin{eqnarray}
\chic^{(\sigma)}(p,m;\tau,-z) &=& \chic^{(\sigma)}(p,-m;\tau,z) ~, ~~~
  (\sigma = \NS, \,\tNS, \, \R)~,\nn
\chic^{(\stR)}(p,m;\tau,-z) &=& -\chic^{(\stR)}(p,-m;\tau,z) ~, ~~~
\label{charge conjugation massive} \\
\chid^{(\sigma)}(s,m;\tau,-z) &=& \chid^{(\sigma)}(N+2K-s,N-m;\tau,z) ~,~~~
 (\sigma = \NS, \,\R, \, \tR) ~, \nn
\chid^{(\stNS)}(s,m;\tau,-z) &=& -\chid^{(\stNS)}(N+2K-s,N-m;\tau,z) ~, 
\label{charge conjugation massless} \\
\chi_0^{(\sigma)}(m;\tau,-z) &=& \chi_0^{(\sigma)}(-m;\tau,z)~ , ~~~
  (\sigma = \NS, \,\tNS, \, \R)~,\nn
\chi_0^{(\stR)}(m;\tau,-z) &=& -\chi_0^{(\stR)}(-m;\tau,z)~ .
\label{charge conjugation graviton} 
\end{eqnarray}

The following identities are also useful $(r\in \bz_N)$;
\begin{eqnarray}
&& \hspace{-1cm}
\chi_0^{(\sigma)}(K(2r+\nu(\sigma));\tau,z)\pm\chid^{(\sigma)}(N,N+K(2r+\nu(\sigma));\tau,z)
+ \chid^{(\sigma)}(2K,K(2r+\nu(\sigma));\tau,z) \nonumber \\
&& \hskip6cm =
\chic^{(\sigma)}\left(i\sqrt{\frac{K}{2N}},K(2r+\nu(\sigma));\tau,z\right)~, 
\label{graviton identity} \\
&&\hspace{-1cm}
\chid^{(\sigma)}(K,m;\tau,z)\pm\chid^{(\sigma)}(N+K,m+N;\tau,z)
  = \chic^{(\sigma)}(p=0,m;\tau,z)~,~~~(\sigma=\NS,\,\R) ~,
\label{boundary identity} 
\end{eqnarray}
where one chooses $+$ ($-$) sign for spin structures NS and R ($\tNS,\tR$). 


Let us next present the modular transformation formulas of the extended 
characters. 
The T-transformation formulas are  quite easy;
\begin{eqnarray}
\chi_*^{(\sigma)}(*,m;\tau+1,z) = e^{2\pi i (h^{(\sigma)}(*,m)-\hat{c}/8)}\,
  \chi_*^{(T\cdot \sigma)}(*,m;\tau,z)~,
\label{T-transformation}
\end{eqnarray}
where the conformal weight $h^{(\sigma)}(*,m)$
can be read off from \eqn{vacua chi c}, \eqn{vacua chi d}, \eqn{vacua
chi 0} and also the formula \eqn{relation NS R}. 
The S-transformation formulas are given in \cite{ES-L}, 
and written as
\begin{eqnarray}
&& \hspace{-1cm}
\chic^{(\sigma)}\left(p,m;-\frac{1}{\tau}, \frac{z}{\tau}\right)
=\kappa(\sigma) e^{i\pi \hat{c}\frac{z^2}{\tau}}
\sqrt{\frac{2}{NK}}
\sum_{m'\in \bsz_{2NK}}\, e^{-2\pi i \frac{m m'}{2NK}}\,
\int_0^{\infty}dp'\, \cos\left(2\pi pp'\right)\,
\chic^{(S\cdot \sigma)}(p',m';\tau,z)~, \nn
&&
\label{S cont} \\
&&\hspace{-1cm}
\chid^{(\sigma)}\left(s,m;-\frac{1}{\tau},\frac{z}{\tau}\right)
=\kappa(\sigma) e^{i\pi \hat{c}\frac{z^2}{\tau}}
\, \left\lb 
\frac{1}{\sqrt{2NK}}\sum_{m'\in \bsz_{2NK}}\, e^{-2\pi i \frac{m m'}{2NK}}
\right.  \nn
&& \hspace{1cm}
  \times \int_0^{\infty} dp'\,
\frac{\cosh\left(2\pi \frac{N-(s-K)}{\sqrt{2NK}}p'\right)
+ e^{2\pi i\left(\frac{m'}{2K}-\frac{1}{2}\nu(S\cdot \sigma)\right)}
\cosh\left(2\pi \frac{s-K}{\sqrt{2NK}}p'\right)}
{2\left|
\cosh \pi \left(\sqrt{\frac{N}{2K}}p'
+i\frac{m'}{2K}-\frac{i}{2}\nu(S\cdot \sigma)\right)
\right|^2} \, \chic^{(S\cdot \sigma)}(p',m';\tau,z) \nn
&& \hspace{1cm}
+ \frac{i}{N} \sum_{s'=K+1}^{N+K-1}\,\sum_{m'\in \bsz_{2NK}}\, 
e^{2\pi i \frac{(s-K)(s'-K)-m m'}{2NK}}\, 
\chid^{(S\cdot \sigma)}(s',m';\tau,z)  \nn
&&  \hspace{1cm} \left.
+ \frac{i}{2N} \sum_{m'\in \bsz_{2NK}}\,
e^{-2\pi i \frac{m m'}{2NK}}\, \left\{
\chid^{(S\cdot \sigma)}(K,m';\tau,z)
- e^{i\pi \nu(\sigma)}\, \chid^{(S\cdot \sigma)}(N+K,N+m';\tau,z)
\right\}
\right\rb~, \nn
&&
\label{S discrete} 
\end{eqnarray}
\begin{eqnarray}
&& \hspace{-1cm}
\chi_0^{(\sigma)}\left(m;-\frac{1}{\tau},\frac{z}{\tau}\right)
= \kappa(\sigma) e^{i\pi \hat{c}\frac{z^2}{\tau}}\, \left\lb
\frac{1}{\sqrt{2NK}}\sum_{m'\in \bsz_{2NK}}\, e^{-2\pi i \frac{m m'}{2NK}}
\right.  \nn
&& \hspace{1cm} \times 
\int_0^{\infty}dp'\, \frac
{\sinh\left(\pi \sqrt{\frac{2K}{N}}p'\right)
\sinh\left(\pi\sqrt{\frac{2N}{K}} p'\right)}
{\left|
\cosh \pi \left(\sqrt{\frac{N}{2K}}p'
+i\frac{m'}{2K}-\frac{i}{2}\nu(S\cdot \sigma)\right)
\right|^2} \, \chic^{(S\cdot \sigma)}(p',m';\tau,z) \nn
&& \hspace{1cm} \left. 
+ \frac{2}{N} \sum_{s'=K+1}^{N+K-1}\, \sum_{m'\in \bsz_{2NK}}\,
\sin \left(\frac{\pi(s'-K)}{N}\right)\, e^{-2\pi i \frac{m m'}{2NK}}\,
\chid^{(S\cdot \sigma)}(s',m';\tau,z)
\right\rb~.
\label{S graviton}
\end{eqnarray}
In these expressions 
$T\cdot \sigma$, $S \cdot \sigma$, $\kappa(\sigma)$ 
and $\nu(\sigma)$ are defined in 
\eqn{T sigma}, \eqn{S sigma},  \eqn{kappa sigma}.
Note that these modular transformation formulas, especially the 
discrete terms appearing in them, are really consistent with 
the formulas of Witten index \eqn{Witten index}. 

It is often useful to extend the S-transformation formula for 
the massive character \eqn{S cont} to the case of continuous 
$U(1)$-charge; 
\begin{eqnarray}
&& \hspace{-1cm}
\chic^{(\sigma)}\left(p,\om;-\frac{1}{\tau}, \frac{z}{\tau}\right)
=\kappa(\sigma) e^{i\pi \hat{c}\frac{z^2}{\tau}}
\sqrt{\frac{2}{NK}}
\sum_{m'\in \bsz_{2NK}}\, 
\int_0^{\infty}dp'\, \cos\left(2\pi pp'\right)\,
\chic^{(S\cdot \sigma)}\left(p',m';\tau,z, \om\right)~,~
 \nn
&&
\label{S cont 2}
\end{eqnarray}
Here the L.H.S is defined by formally replacing the discrete parameter
$m\in \bz_{2NK}$ with a continuous one $0\leq \om < 2NK$ in \eqn{chi c}.
In the R.H.S we introduced 
\begin{eqnarray}
&& \chic^{(\sigma)}(p,m;\tau,z,w) = q^{\frac{p^2}{2}} \Th{m}{NK}
\left(\tau,\frac{2z}{N}- \frac{w}{NK}\right) \,
\frac{\th_{\lb \sigma \rb}(\tau,z)}{\eta(\tau)^3}~, 
\label{extended massive w}  
\end{eqnarray}
where we set $\th_{\lb \sigma \rb}= \th_3, \th_4, \th_2, i\th_1$
for $\sigma = \NS, \tNS, \R, \tR$. 
$\chic^{(\sigma)}(p,m;\tau,z,w)$ is the extension of extended 
massive character with one parameter $0\leq w < 2NK$
that characterizes the relative phases of irreducible characters 
in the sum over spectral flow. Note that the R.H.S in \eqn{S cont 2}
contains only contributions with discrete $U(1)$-charges.


~


\section{Twisted $\cN=2$ Characters}

~

The twisted $\cN=2$ characters are defined by the twisting
with respect to  a $\bz_2$-automorphism in the $\cN=2$ SCA;
\begin{equation}
\sigma ~:~ T\,\longrightarrow\, T,~~~J\,\longrightarrow\, -J,~~~
G^{\pm}\,\longrightarrow\, G^{\mp} ~. 
\label{sigma twist}
\end{equation}
We denote the twisted characters as $\ch{(*)}{(S,T)}$ where
$*$ expresses the spin structure and $S,T=\pm$ express the spatial 
and temporal boundary conditions of the $\sigma$-twist. 
The relevant quantum number is only the conformal weight, since 
the $\sigma$-twisting leaves only the states with vanishing
$U(1)$-charge.  
It is easy to see the following identities (see {\em e.g.}
\cite{ES-G2orb});
\begin{eqnarray}
&& \ch{(\sNS)}{(+,-)}(\tau)= \ch{(\stNS)}{(+,-)}(\tau)~, ~~~
\ch{(\sNS)}{(-,+)}(\tau) = \ch{(\sR)}{(-,+)}(\tau)~, 
~~~ \ch{(\stNS)}{(-,-)}(\tau)= \ch{(\sR)}{(-,-)}(\tau) ~, 
\label{group 1} \\
&&  \ch{(\sR)}{(+,-)}(\tau)= \ch{(\stR)}{(+,-)}(\tau)~, ~~~
\ch{(\stNS)}{(-,+)}(\tau) = \ch{(\stR)}{(-,+)}(\tau)~, 
~~~ \ch{(\sNS)}{(-,-)}(\tau)= \ch{(\stR)}{(-,-)}(\tau) ~ .
\label{group 2}
\end{eqnarray}
All the characters in the second line \eqn{group 2} actually vanish
due to fermion zero modes, and we are left only three non-trivial 
twisted characters \eqn{group 1}, 
which we denote as $\chi_{(+,-)}(\tau)$, $\chi_{(-,+)}(\tau)$ and 
$\chi_{(-,-)}(\tau)$.

The twisted massive characters in any $\cN=2$ SCFT's with $\hat{c}>1$
are essentially trivial. We can readily calculate them as
\begin{eqnarray}
&&\chi_{(+,-)}(p;\tau) = \frac{2 q^{\frac{p^2}{2}}}{\th_2(\tau)}~, ~~~
(h= \frac{p^2}{2} + \frac{\hat{c}-1}{8})~, \nn
&&\chi_{(-,+)}(p;\tau) = \frac{2 q^{\frac{p^2}{2}}}{\th_4(\tau)}~, ~~~
(h= \frac{p^2}{2} + \frac{\hat{c}}{8})~, \nn
&&\chi_{(-,-)}(p;\tau) = \frac{2 q^{\frac{p^2}{2}}}{\th_3(\tau)}~, ~~~
(h= \frac{p^2}{2} + \frac{\hat{c}}{8})~. 
\label{twisted massive}
\end{eqnarray}

In the minimal model $M_{k}$ the twisted characters are much more
involved. The relevant formulas are summarized in
\cite{ES-G2orb} based on the results \cite{Dobrev,ZF2,Qiu};
\begin{eqnarray}
\chi_{\ell \,(+,-)}(\tau)& =& 
\left\{
\begin{array}{ll}
\dsp  \frac{2}{\th_2(\tau)} \left(
\Th{2(\ell+1)}{4(k+2)}(\tau)+(-1)^k\Th{2(\ell+1)+4(k+2)}{4(k+2)}(\tau)
\right)   &  ~~ (\ell~:~\mbox{even}) ,\\
0 & ~~(\ell~:~\mbox{odd}).
\end{array}
\right.
\nn
\chi_{\ell\,(-,+)}(\tau)&=& \frac{1}{\th_4(\tau)}
\,\left(\Th{\ell+1-\frac{k+2}{2}}{k+2}(\tau)-
\Th{-(\ell+1)-\frac{k+2}{2}}{k+2}(\tau)\right)  \nn
&= & \frac{1}{\th_4(\tau)}\left(\Th{2(\ell+1)-(k+2)}{4(k+2)}(\tau)
 + \Th{2(\ell+1)+3(k+2)}{4(k+2)}(\tau) \right. \nn
 && \left. - \Th{-2(\ell+1)-(k+2)}{4(k+2)}(\tau)
 - \Th{-2(\ell+1)+3(k+2)}{4(k+2)}(\tau) \right)~,
\nn
\chi_{\ell\,(-,-)}(\tau)&=& 
\frac{1}{\th_3(\tau)}
\left(\Th{2(\ell+1)-(k+2)}{4(k+2)}(\tau)
 +(-1)^k \Th{2(\ell+1)+3(k+2)}{4(k+2)}(\tau) \right. \nonumber \\
 && \left. +(-1)^{\ell} \Th{-2(\ell+1)-(k+2)}{4(k+2)}(\tau)
 +(-1)^{k+\ell} \Th{-2(\ell+1)+3(k+2)}{4(k+2)}(\tau) \right)~.
\label{twisted minimal}
\end{eqnarray}
The first character has the vacuum with 
\begin{eqnarray}
h= h_{\ell} \equiv \frac{\ell(\ell+2)}{4(k+2)}~, ~~~(\mbox{the same as
the primary of $SU(2)_k$})~,
\end{eqnarray}
and, the second and third ones have the vacuum with 
\begin{eqnarray}
h= h_{\ell}^{t} \equiv  
\frac{k-2+(k-2\ell)^2}{16(k+2)}+\frac{1}{16}~.
\label{h t}
\end{eqnarray}
This vacuum of \eqn{h t} is interpreted as given by the product of 
the twist field in the $U(1)$-sector and 
the ``$C$-disorder field'' \cite{ZF2} in the $\bz_k$-parafermion  
theory \cite{ZF1}. Note that we have $\chi_{k-\ell\,(-,+)}=
\chi_{\ell\,(-,+)}$,  $\chi_{k-\ell\,(-,-)}=
\chi_{\ell\,(-,-)}$.  
We have to actually identify the corresponding primary fields, 
only leaving 
$\ell=0, 1,\ldots, \left\lb \frac{k}{2}\right\rb$ as 
independent primary fields.

The modular transformation formulas are written as 
\begin{eqnarray}
&&\hskip-15mm
\chi_{\ell\,(+,-)}(\tau+1)= 
e^{2\pi i \left(h_\ell -\frac{k}{8(k+2)}\right)}\,
\chi_{\ell\,(+,-)}(\tau)~, 
\hskip5mm
\chi_{\ell\,(+,-)}\left(-\frac{1}{\tau}\right)
=\sum_{\ell'=0}^k\, (-1)^{\ell/2} 
S_{\ell,\ell'}\,
\chi_{\ell'\,(-,+)}(\tau),  \nn
&&\hskip-15mm
\chi_{\ell\,(-,+)}(\tau+1)=
e^{2\pi i\left(h^t_\ell-\frac{k}{8(k+2)}\right)}\, 
\chi_{\ell\,(-,-)}(\tau)~, 
\hskip5mm
\chi_{\ell\,(-,+)}\left(-\frac{1}{\tau}\right)=
\sum_{\ell'=0}^k\, S_{\ell,\ell'}(-1)^{\ell'/2}\, 
\chi_{\ell'\,(+,-)}(\tau)~, \nn
&&\hskip-15mm \chi_{\ell\,(-,-)}(\tau+1)= 
e^{2\pi i \left(h^t_\ell -\frac{k}{8(k+2)}\right)}\,
\chi_{\ell\,(-,+)}(\tau)  ~, 
\hskip5mm
\chi_{\ell\,(-,-)}\left(-\frac{1}{\tau}\right)=(-i)\,
\sum_{\ell'=0}^k\, \widehat{S}_{\ell,\ell'}\,
\chi_{\ell'\,(-,-)}(\tau) ~ .
\label{modular twisted minimal} 
\end{eqnarray}
Here 
$ S_{\ell,\ell'}\equiv \sqrt{\frac{2}{k+2}}
\sin\left(\frac{(\ell+1)(\ell'+1)}{k+2}\right)$ is 
the modular S-matrix of the $SU(2)$ WZW model at level $k$, 
and  we set $\widehat{S}_{\ell,\ell'}= 
e^{\frac{\pi i}{2}\left(\ell+\ell'+2-\frac{k+2}{2}\right)}\, 
S_{\ell,\ell'}$.

~


\section{Extended Characters and Appell Function}


~

In \cite{ES-L} the modular transformation formulas of the massless 
extended characters in the $L_{N,K}$-sector
have been derived using the integration formula 
presented in \cite{Miki}. On the other hand, the massless extended
characters are closely related with the higher level Appell functions 
\cite{Pol,STT}, and their modular properties are studied in \cite{STT}.  
In this comparably long Appendix 
we try to rederive the modular transformation formulas
of massless extended characters based on the results given in \cite{STT}.
We shall use the notations $e^{2\pi i \tau} \equiv q$, 
$e^{-2\pi i \frac{1}{\tau}} \equiv \tq$ and $e^{2\pi i z}\equiv y$. 
For our purpose it is the most convenient to concentrate on the $\tR$-sector.

~

\subsection{Preliminaries}

~

The relevant modular transformation formula \eqn{S discrete}
can be rewritten in a more convenient form (for the $\tR$-sector); 
\begin{eqnarray}
&&\hspace{-1cm}
\chid^{(\stR)}\left(s,m;-\frac{1}{\tau},\frac{z}{\tau}\right)
= e^{i\pi \hat{c}\frac{z^2}{\tau}}
\, \left\lb 
-\frac{i}{\sqrt{2NK}}
\sum_{m'\in \bsz_{2NK}}\, e^{-2\pi i \frac{m m'}{2NK}} \,
 \right.   
\nn
&& \hspace{1cm} \left. \times 
  \int_{\bsr+i0} dp'\,
\frac{e^{-2\pi \frac{s-K}{\sqrt{2NK}}p'}}
 {1-e^{-2\pi \left(\sqrt{\frac{N}{2K}}p'+i\frac{m'}{2K}\right)}}\,
\chic^{(\stR)}(p',m';\tau,z)  \right. \nn
&& \hspace{1cm} 
\left.
+ \frac{1}{N} \sum_{s'=K}^{N+K-1}\,\sum_{m'\in \bsz_{2NK}}\, 
e^{2\pi i \frac{(s-K)(s'-K)-m m'}{2NK}}\, 
\chid^{(\stR)}(s',m';\tau,z) 
\right\rb~, 
\label{S discrete 2} 
\end{eqnarray}
The extended massive character $\chic^{(\stR)}(p,m)$ is explicitly written 
as
\begin{eqnarray}
\chic^{(\stR)}(p,m;\tau,z) \equiv q^{\frac{p^2}{2}} 
\Th{m}{NK}\left(\tau,\frac{2z}{N}\right)\, \frac{i\th_1(\tau,z)}
{\eta(\tau)^3}~.
\end{eqnarray}


On the other hand, the modular transformation formula
of the level $\ell$ Appell function \eqn{Appell}
is given by \cite{STT};
\begin{eqnarray}
{\cal K}_{\ell}\left(-{1\over \tau},{\nu\over \tau},{\mu\over \tau}\right)
&=&\tau e^{i\pi\ell{\nu^2-\mu^2\over\tau}}
{\cal K}_{\ell}(\tau,\nu,\mu)
+\tau\sum_{a=0}^{\ell-1}\,e^{i\pi{\ell\over \tau}
(\nu+{a \over \ell}\tau)^2}
\Phi(\ell\tau,\ell\mu-a\tau)\theta_3(\ell\tau,\ell\nu+a\tau) \nn
&\equiv & \tau e^{i\pi\ell{\nu^2-\mu^2\over\tau}}
{\cal K}_{\ell}(\tau,\nu,\mu)
+\tau e^{i\pi \ell \frac{\nu^2}{\tau}}\sum_{a=0}^{\ell-1}\,
\Phi(\ell\tau,\ell\mu-a\tau) \Th{a}{\frac{\ell}{2}}\left(\tau,2\nu\right) ~,
\label{STT formula}
\end{eqnarray}
where we set
\begin{equation}
\Phi(\tau,\mu)\equiv 
-{i\over 2\sqrt{-i\tau}}-{1\over 2}\int_{-\infty}^{\infty} dx \, 
e^{-\pi x^2}
{\sinh(\pi x\sqrt{-i\tau}(1+2{\mu\over \tau}))\over\sinh(\pi x\sqrt{-i\tau})}
~.
\label{Phi}
\end{equation}
As we see below, it is enough to only consider 
the $\mu=0$ case.
Using the identity 
\begin{eqnarray}
\frac{\tq^{\frac{x'^2}{2}}}{\sqrt{-i\tau}} 
= \int_{\bsr+i\xi} dp'\, e^{-2\pi i p' x'} q^{\frac{p'^2}{2}}~,~~~
({}^{\forall} \xi \in \br)~,
\end{eqnarray}
and also the standard contour 
deformation technique, we can rewrite  \eqn{STT formula}
in a more convenient form;
\begin{eqnarray}
{\cal K}_{\ell}\left(-{1\over \tau},{\nu\over \tau},0\right)
&= & \tau e^{i\pi\ell{\nu^2\over\tau}} \, \left\lb 
{\cal K}_{\ell}(\tau,\nu,0)
-\frac{i}{\sqrt{\ell}} \sum_{a=0}^{\ell-1}\,
\Th{a}{\frac{\ell}{2}}\left(\tau,2\nu\right)
\int_{\bsr+i0}dp'\, \frac{1}{1-e^{-2\pi 
\left(\frac{p'}{\sqrt{\ell}}+i\frac{a}{\ell}\right)}}\, 
q^{\frac{p'^2}{2}} \right\rb ~. \nn
&&
\label{STT formula 3}
\end{eqnarray}

~


\subsection{Derivation of \eqn{S discrete 2} from the STT 
Formula \eqn{STT formula 3}}

~

Now, we try to derive the modular transformation formula 
\eqn{S discrete 2} from \eqn{STT formula 3}. 
The key idea is to make use of the relations between 
the massless characters and Appell function \eqn{relation chid cK}, 
\eqn{relation chid cK 2}.

We first focus on the range $K \leq  s \leq N+K-1$, and 
later discuss other values. 
Based on \eqn{relation chid cK 2},
our first task is to evaluate the modular transformation of
$\dsp \cK_{2NK}\left(\tau, \frac{z+a\tau+b}{N},0\right)$ by the formula
\eqn{STT formula 3} $(a,b \in \bz_N)$;
\begin{eqnarray}
&&\hspace{-1cm}
\cK_{2NK}\left(-\frac{1}{\tau},\frac{1}{N}
\left(\frac{z}{\tau}-\frac{a}{\tau}+b\right), 0\right)
\equiv \cK_{2NK}\left(-\frac{1}{\tau},\frac{1}{\tau}
\left(\frac{z+b\tau-a}{N}\right), 0\right) \nn
&& \hspace{5mm}
= \tau e^{i\pi 2NK \frac{1}{\tau}\left(\frac{z+b\tau-a}{N}\right)^2}\,
\left\lb 
\cK_{2NK}\left(\tau,\frac{z+b\tau-a}{N},0\right) \right. \nn
&& \hspace{5mm}
\left.
 -\frac{i}{\sqrt{2NK}} \sum_{m'\in \bsz_{2NK}}\, \Th{m'}{NK}\left(
  \tau, \frac{2(z+b\tau-a)}{N}\right) 
\, \int_{\bsr+i0}dp'\, \frac{1}{1-e^{-2\pi \left( 
\frac{p'}{\sqrt{2NK}}+i\frac{m'}{2NK}\right)}}\, q^{\frac{p'^2}{2}}
\right\rb ~.\nn
&& 
\label{derivation 1}
\end{eqnarray}


Using  \eqn{relation chid cK 2},  we obtain 
\begin{eqnarray}
&& \hspace{-1cm}
\chid^{(\stNS)}\left(s,s-K+2Ka;-\frac{1}{\tau},\frac{z}{\tau}\right) \nn
&& \hspace{5mm}
= \tq^{\frac{K}{N}a^2} e^{2\pi i \frac{2K}{N} \frac{az}{\tau}}\,
\frac{1}{N} \sum_{b\in \bsz_N}\, e^{-2\pi i \frac{(s-K)b}{N}}\,
\cK_{2NK}\left(-\frac{1}{\tau}, \frac{1}{N}
\left(\frac{z}{\tau}-\frac{a}{\tau}+b\right), 0\right)\, 
\frac{i\th_1(-1/\tau,z/\tau)}{\eta(-1/\tau)^3} \nn
&& \hspace{5mm} = e^{i\pi \hat{c}\frac{z^2}{\tau}} 
\frac{1}{N} \sum_{b\in \bsz_N}\, q^{\frac{K}{N}b^2}y^{\frac{2K}{N}b}
e^{-2\pi i \frac{s-K+2Ka}{N}b}\,  \left\lb
\cK_{2NK}\left(\tau,\frac{z+b\tau-a}{N},0\right)\,
 \right. \nn
&& \hspace{5mm} \left.  -\frac{i}{\sqrt{2NK}} \sum_{m'\in \bsz_{2NK}}\,
\Th{m'}{NK}\left(\tau, \frac{2(z+b\tau-a)}{N}\right)\,
\int_{\bsr+i0}dp'\, \frac{1}{1-e^{-2\pi 
\left(\frac{p'}{\sqrt{2NK}}+i\frac{m'}{2NK}\right)}}\, q^{\frac{p'^2}{2}}
\right\rb \,  \frac{i\th_1(\tau,z)}{\eta(\tau)^3}~. \nn
&&\hspace{5mm}
\equiv \mbox{\bf (dis. term)} + \mbox{\bf (con. term)}~.
\label{derivation 2} 
\end{eqnarray}
The terms {\bf (dis. term)} and {\bf (con. term)}
correspond to the first and second terms, 
which we will separately evaluate.

~

\noindent
{\bf 1. discrete term : }

Recalling the relation \eqn{relation chid cK}, 
we obtain
\begin{eqnarray}
\mbox{\bf (dis. term)} &=& e^{i\pi \hat{c} \frac{z^2}{\tau}}\,
\frac{1}{N} \sum_{b\in \bsz_N}\, \sum_{s'=K}^{N+K-1}\,
e^{-\frac{2\pi i }{N}\left\{(s-K)b+(s'-K)a+2Kab\right\}}\,
\chid^{(\stR)}(s',s'-K+2Kb;\tau,z) \nn
&=& e^{i\pi \hat{c} \frac{z^2}{\tau}}\,
\frac{1}{N} \sum_{m'\in \bsz_{2NK}}\, \sum_{s'=K}^{N+K-1}\,
e^{2\pi i \frac{(s-K)(s'-K)-mm'}{2NK}}\, \chid^{(\stR)}(s',m';\tau,z)~.
\label{derivation dis}
\end{eqnarray}
In the second line we set $m=s-K+2Ka$, $m'=s'-K+2Kb$.

~

\noindent
{\bf 2. continuous term : }

We first note 
\begin{eqnarray}
&&q^{\frac{K}{N}b^2}y^{\frac{2K}{N}b}\, \Th{m'}{NK}\left
(\tau, \frac{2(z+b\tau-a)}{N}\right) =
e^{-2\pi i \frac{a m'}{N}}\, \Th{m'+2Kb}{NK}\left(\tau,\frac{2z}{N}\right)~,
\end{eqnarray}
Then we obtain 
\begin{eqnarray}
\hspace{-1cm} \mbox{\bf (con. term)} 
&=& 
 e^{i\pi \hat{c}\frac{z^2}{\tau}} \frac{1}{N}
\sum_{b\in \bsz_N}\, \sum_{m'\in \bsz_{2NK}}\, e^{-2\pi i \frac{(s-K)b}{N}}
e^{-2\pi i \frac{am'}{N}}\, 
\Th{m'}{NK}\left(\tau,\frac{2z}{N}\right) \nn
&& \hspace{2cm} \times 
\frac{-i}{\sqrt{2NK}} \int_{\bsr+i0}dp' \,q^{\frac{p'^2}{2}}\,
\frac{1}{1-e^{-2\pi \left(\frac{p'}{\sqrt{2NK}}
+i\frac{m'-2Kb}{2NK}\right)}}\,
\frac{i\th_1(\tau,z)}{\eta(\tau)^3}~.
\label{derivation con 1}
\end{eqnarray}
Moreover, using the identities
\begin{eqnarray}
&& 
\frac{1}{1-e^{-2\pi 
\left(\frac{p'}{\sqrt{2NK}}+i\frac{m'-2Kb}{2NK}\right)}} =
\sum_{\al=0}^{N-1} \, 
\frac{e^{-2\pi \al\left(\frac{p'}{\sqrt{2NK}}+i\frac{m'-2Kb}{2NK}\right)}}
{1-e^{-2\pi \left(\sqrt{\frac{N}{2K}}p'+i\frac{m'}{2K}\right)}}~, \nn
&&
\frac{1}{N}\sum_{b\in \bsz_N}\,e^{-2\pi i \frac{b}{N}(s-K-\al)}
= \delta_{\al,s-K}^{(N)}~,
\end{eqnarray}
we find 
\begin{eqnarray}
\hspace{-5mm}
\mbox{\bf (con. term)} &=& -e^{i\pi \hat{c}\frac{z^2}{\tau}} 
\frac{i}{\sqrt{2NK}}
\sum_{m'\in \bsz_{2NK}}\, e^{-2\pi i \frac{m m'}{2NK}}\, 
\Th{m'}{NK}\left(\tau,\frac{2z}{N}\right) \, \nn
&& \hspace{2cm} \times
\int_{\bsr+i0}dp'\, q^{\frac{p'^2}{2}}\, 
\frac{e^{-2\pi \frac{s-K}{\sqrt{2NK}}p'}}
{1-e^{-2\pi \left( \sqrt{\frac{N}{2K}}p'+i\frac{m'}{2K}\right)}}\,
\frac{i\th_1(\tau,z)}{\eta(\tau)^3}~. \nn
&=& - e^{i\pi \hat{c}\frac{z^2}{\tau}} \frac{i}{\sqrt{2NK}}
\sum_{m'\in \bsz_{2NK}}\, e^{-2\pi i \frac{m m'}{2NK}}\,
\int_{\bsr+i0}dp'\,  \frac{e^{-2\pi \frac{s-K}{\sqrt{2NK}}p'}}
{1-e^{-2\pi \left(\sqrt{\frac{N}{2K}}p'+i\frac{m'}{2K}\right)}}\, 
\chic^{(\stR)}(p',m';\tau,z)~. \nn
&& \label{derivation con 2}
\end{eqnarray}
In this way we have reproduced the modular transformation
formula \eqn{S discrete 2}.

To complete our proof we have to also work with
the cases of $s<K$ and $N+K\leq s$.
First of all, $s=N+K$ is reduced to $s=K$ by the charge conjugation
relation \eqn{charge conjugation massless}. 
Next we consider the case of $s>N+K$. We set 
\begin{eqnarray}
s-K=s_0-K +Nj~,~~~ 0\leq s_0-K \leq N~,~~~ j\in \bz_{>0}~,
\label{s0}
\end{eqnarray} 
and 
\begin{eqnarray}
m=s-K+2Kr~,~~~ m_0=s_0-K+2Kr~,~~~(i.e.~ m=m_0+Nj)~.
\label{m0}
\end{eqnarray}
Making use of the identity
\begin{eqnarray}
\frac{\left(yq^{N\left(n+\frac{r}{N}\right)}\right)^{\frac{s_0-K}{N}+j}}
{1-yq^{N\left(n+\frac{r}{N}\right)}}
= \frac{\left(yq^{N\left(n+\frac{r}{N}\right)}\right)^{\frac{s_0-K}{N}}}
{1-yq^{N\left(n+\frac{r}{N}\right)}}
-\sum_{a=0}^{j-1}\, 
\left(yq^{N\left(n+\frac{r}{N}\right)}\right)^{\frac{s_0-K}{N}+a}~,
\label{identity 1}
\end{eqnarray}
we can show
\begin{eqnarray}
\chid^{(\stR)}(s,m;\tau,z)= \chid^{(\stR)}(s_0,m_0;\tau,z)
-\sum_{a=0}^{j-1}\chic^{(\stR)}\left(i\frac{s_0-K+Na}{\sqrt{2NK}}, 
m_0+Na;\tau,z
\right)~.
\label{identity 2}
\end{eqnarray}
Since $K\leq s_0 \leq N+K$, 
the first term $\chid^{(\stR)}(s_0,m_0;\tau,z)$
has been already shown to obey the modular transformation formula 
\eqn{S discrete 2}, 
and the modular transformation of second term is evaluated as
\begin{eqnarray}
&& \hspace{-1cm}
 \chic^{(\stR)}\left(i\frac{s_0-K+Na}{\sqrt{2NK}}, 
m_0+Na ; -\frac{1}{\tau},
 \frac{z}{\tau}\right) \nn
&& =
 - e^{i\pi \hat{c}\frac{z^2}{\tau}}
\frac{i}{\sqrt{2NK}} \sum_{m'\in \bsz_{2NK}}\,
e^{-2\pi i \frac{(m_0+Na)m'}{2NK}}\, 
 \int_{-\infty}^{\infty}dp'\,
 e^{-2\pi \frac{s_0-K+Na}{\sqrt{2NK}}p'}\,\chic^{(\stR)}(p',m';\tau,z)~.
\label{identity 3}
\end{eqnarray}
We also remark
\begin{eqnarray}
&& \hspace{-1cm}\frac{e^{-2\pi \left(\sqrt{\frac{N}{2K}}p'
+i\frac{m'}{2K}\right)\frac{s_0-K}{N}}}
{1-e^{-2\pi \left(\sqrt{\frac{N}{2K}}p'+i\frac{m'}{2K}\right)}}
-\sum_{a=0}^{j-1}\, e^{-2\pi \left(\sqrt{\frac{N}{2K}}p'+i\frac{m'}{2K}\right)
\left(\frac{s_0-K}{N}+a\right)} = 
\frac{e^{-2\pi \left(\sqrt{\frac{N}{2K}}p'+i\frac{m'}{2K}\right)
\left(\frac{s_0-K}{N}+j\right)}}
{1-e^{-2\pi \left(\sqrt{\frac{N}{2K}}p'+i\frac{m'}{2K}\right)}}~,
\label{identity 4} \\
&& \hspace{-1cm} e^{2\pi i \frac{(s-K)(s'-K)-mm'}{2NK}}
= e^{2\pi i \frac{(s_0-K)(s'-K)-m_0 m'}{2NK}}~.
\label{identity 5}
\end{eqnarray}
Combining these identities with the character relation 
\eqn{identity 2}, we can readily confirm that 
$\chid^{(\stR)}(s,m;\tau,z)$ with $s>N+K$
still obeys the modular transformation formula \eqn{S discrete 2}.

The cases of $s<K$ can be reduced to 
$s>N+K$ by using again  the charge conjugation relation \eqn{charge
conjugation massless}.
Therefore, the formula \eqn{S discrete 2} has been derived from
\eqn{STT formula 3} in all the cases.


\end{document}